# Multi-objective Evolutionary Algorithms (MOEAs) in PMEDM - A Comparative Study in Pareto Frontier

**Mohsen Asghari Ilani[1], Yaser Mike Banad[1]**

[1] School of Electrical and Computer Engineering, University of Oklahoma, Norman, 73019, U.S.A.

## Abstract

Electrical discharge machining (EDM) is a crucial process in precision manufacturing, leveraging electro-thermal energy to remove material without electrode contact. In this study, we delve into the realm of Machine Learning (ML) to enhance the efficiency and precision of EDM, particularly focusing on Powder-Mixed Electrical Discharge Machining (PMEDM) with the integration of a vibration system. We comprehensively evaluate four leading ML models – Deep Neural Network (DNN), Extreme Gradient Boosting (XGBoost), Adaptive Gradient Boosting (AdaBoost), and ElasticNet - against a pool of ML models, employing various evaluation metrics including Accuracy ($R2$), Mean Squared Error (MSE), Root Mean Squared Error (RMSE), and Mean Absolute Error (MAE). Our evaluations, conducted on datasets enriched with features derived from powder addition and electrode vibration, reveal XGBoost's superior accuracy, followed by AdaBoost, DNN, and ElasticNet. Furthermore, through the integration of Multi-Objective Evolutionary Algorithms (MOEAs) such as NSGA-II, NSGA-III, UNSGA-III, and C-TAEA, we explore and optimize the Pareto front to attain optimal solutions. Our findings underscore the transformative potential of ML and optimization techniques in advancing EDM processes, offering cost-effective, time-efficient, and reliable solutions for precision manufacturing applications.

## 1. Introduction

Electro-Discharge Machining (EDM) has emerged as an advanced manufacturing technique designed to address shortcomings of conventional methods by accommodating complex geometries and high material hardness, while simultaneously reducing material consumption and production lead times. Its ability to enhance performance, reliability, and customization has positioned EDM as a powerful tool across macro, micro, and nano-scale industries, particularly in fields like bio-applications and aerospace technologies (Joshi & Pande, 2009; Wang et al., 2003). In response to evolving industry needs, researchers and practitioners are increasingly exploring enhancements to EDM setups. These include innovations such as Powder Mixed-EDM (PMEDM), which involves introducing powders into the machining gas to control ionizing plasma channels, and systems incorporating electrode vibration. These advancements find application in nano-welding, machining, and cutting processes, offering precise material geometry, customizable compositions, and improved defect detection while transitioning from conventional to advanced manufacturing practices. In PMEDM, the integration of vibration systems facilitates electro-thermal energy production without direct electrode contact, relying on plasma channels formed within the discharge gap (Ilani & Khoshnevisan, 2022). However, challenges persist in material removal, as uncontrolled electrode wear rates (EWR) and material removal rates (MRR) can adversely affect part structural integrity and surface finish. Overcoming these challenges remains a complex task, given the multi-scale physics involved in the PMEDM process.

To ensure better control and stability in machining operations, it's vital to closely monitor key parameters both in real-time (in-situ monitoring) and after machining (ex-situ monitoring). Parameters such as Material Removal Rate (MRR), Electrode Wear Rate (EWR), and surface roughness (Ra) are critical in preventing defects that could compromise the integrity of the finished parts. However, overcoming these challenges remains highly complex due to the intricate physics involved in EDM, demanding precise control and management. Monitoring techniques offer valuable insights into enhancing the control of PMEDM setups, especially with the addition of electrode vibration systems and the introduction of powders into the machining gap (Ilani & Khoshnevisan, 2021). These methods aid in identifying and addressing common defects that arise during the EDM process, focusing on factors affecting the Heat-Affected Zone (HAZ), melt pool behavior, and the evaporation process.

Moreover, these monitoring techniques indirectly affect the material removal rate from the electrodes, which is crucial for maintaining surface quality during post-processing. Techniques like thermography, which visualize heat distribution within the melt pool, assist in monitoring phase transitions, microstructure alterations, and the formation of new phases. Common defects in EDM, such as spatter, cracks, and porosity, often stem from the dynamics of the melt pool and the complex processes of heat conduction, radiation, and convection within the EDM gap. Hence, monitoring these processes is indispensable for defect mitigation. Parameters such as MRR, EWR, and Ra serve as standard metrics for evaluating the EDM process, irrespective of whether powders are added or electrodes are vibrated. Analyzing these parameters provides researchers and practitioners with a comprehensive understanding of the EDM process and its intricacies, aiding in process optimization and quality control.

The complexity of PMEDM processes stems from various factors, including the dielectric coefficient before and after adding powders, properties of electrodes and powders, electrical discharges within the machining gap, thermal effects on the melt pool and HAZ, as well as symmetrical and asymmetrical melt pool growth. Additionally, changes in material properties and microstructure during heating, melting, evaporating, and cooling phases contribute to the intricacies of PMEDM processes. The combination of physical, thermal, material, and optical features has made PMEDM processes highly complex, with no single study considering all aspects of the problem. This multi-physics and multi-scale nature of EDM processes, coupled with the significant influence of processing parameters on the quality of machined products, has prompted a shift in PMEDM approaches. Traditionally, PMEDM relied solely on physics-based methodologies. However, there has been a transition towards combining physics-based and data-driven approaches. Datasets extracted primarily focus on melt pool characteristics, given their critical role in defect formation in PMEDM-built products.

Consequently, data-driven analysis and ML have become integral to advanced manufacturing applications, particularly in EDM and PMEDM research. While experimental in-situ monitoring techniques are effective, they can be costly, inefficient, and require extensive preparation and calibration. Alternatively, employing ML models built on experimental data offers a more cost-effective solution. With a reliable training dataset, ML models can make accurate predictions and efficiently determine optimal processing parameters, benefiting future EDM setups. By utilizing ML methods in EDM and PMEDM, manufacturers can enhance control and effectively monitor defect formation processes, optimize manufacturing parameters, and reduce costs. Moreover, compared to experimental techniques tailored for specific subjects, ML models offer greater flexibility and can easily adapt to various applications. This facilitates the identification of meaningful correlations between process parameters and mechanisms of EDM defect formation.

Adapting PMEDM processes to varying machine configurations, adjusting process parameters, and incorporating different materials and powder additions poses a significant challenge in establishing reliable datasets or models for widespread application. Compounding this challenge is the time-consuming and costly nature of acquiring data specific to PMEDM operations. These limitations on data availability in PMEDM persist despite assuming that calibrated sensors accurately record measurements. Furthermore, selecting the appropriate algorithm based on input features is crucial in developing models that accurately capture the behaviors of input-target relationships. Given that PMEDM processes involve multiple physics phenomena, directly and indirectly, influencing defect formation, such as melt pool geometry, microstructure, material properties, and the dynamics of the melt pool during heating, melting, and cooling phases, establishing meaningful relationships becomes inherently complex. This complexity extends to identifying linear, non-linear, or combined relationships among these factors. Consequently, creating machine learning models for additive manufacturing presents challenges due to the scarcity of data and the intricate nature of PMEDM processes.

On the other hand, the concept of the Pareto frontier, often referred to as the Multi-objective Evolutionary (MOE) theory, originated from the observations of Italian economist Vilfredo Pareto. He noted that in his garden, roughly 20% of pea pods produced approximately 80% of the healthy peas. This idea of weighting and monitoring the effects of inputs on outputs has gained popularity. In the case of thermo-electrical processes like EDM, which involve multiple physics, achieving a meaningful strategy to obtain accurate results is challenging and complex. Understanding the proportional effectiveness of inputs and their impact on outputs is crucial. While this strategy works well for machine learning algorithms, the complexity of the process can sometimes lead to decreased prediction accuracy. Additionally, collecting datasets for EDM processes presents challenges due to the multi-scale nature of process parameters, which is time-consuming and costly. It's important to note that a significant portion of results often stems from a minority of causes. Factors such as the type of machine, addition of powders, maximum voltage, and current applied, materials and their composition, as well as physical properties, including electrical and thermal conductivity coefficients, density, melting point, and boiling point, play crucial roles in EDM processes. Other parameters like plasma channel and ionization also contribute to the overall outcome. Researchers and practitioners in various industrial applications evaluate EDM processes based on three main outputs: MRR, EWR, and Ra. These outputs provide valuable insights into the effectiveness of EDM processes and guide decision-making in engineering applications. Furthermore, MOE theory, in conjunction with ML and considering the limitations of available datasets, plays a crucial role in making informed decisions regarding the conditions applied to the output. Typically, the evaluation of EDM processes aims for a higher MRR, lower EWR, and reduced Ra simultaneously while ensuring that the surface integrity and material properties fall within the desired range of specifications(Mert Doleker et al., 2019; Zhang et al., 2010).

In the broad application of advanced manufacturing, particularly in EDM, nature-inspired optimization algorithms have surpassed deterministic approaches, particularly in tackling single and multi-objective optimization challenges. These algorithms, drawing inspiration from natural phenomena, include Particle Swarm Optimization (PSO), Genetic Algorithms (GA) (Zhang et al., 2010), Ant Colony Optimization (ACO), and Artificial Bee Colony Optimization (ABCO). They are widely embraced for predicting machining performance metrics and conducting parametric optimization, especially in scenarios where multiple factors like process dynamics, material properties, and environmental conditions must be considered.

However, achieving optimal outcomes with these optimization methods requires careful adjustment of various algorithm parameters. Neglecting this step can lead to suboptimal results in achieving global optimization objectives. For example, Bandhu et al. (Bandhu et al., 2021) effectively utilized NSGA-II to optimize welding parameters, obtaining a Pareto front for optimal conditions. Similarly, Neeraj Sharma et al. (N. Sharma et al., 2019) applied the Taguchi gray relation approach to optimize processing parameters for Ti6Al4V using Wire Electrical Discharge Machining (WEDM). Devarasiddappa et al. (Devarasiddappa & Chandrasekaran, 2020) utilized a modified teaching–learning algorithm to optimize minimum surface roughness for titanium alloys, showcasing its accuracy and consistency. Priyaranjan Sharma et al. (P. Sharma et al., 2021) combined Teaching–Learning-Based Optimization (TLBO) (Devarasiddappa & Chandrasekaran, 2020) with Response Surface Methodology (RSM) to derive a Pareto front for surface roughness and material removal rate for alloy 706. TLBO, simulating the teaching–learning process in a classroom, involves an instructor phase and a learner phase, with no algorithm-specific parameters and straightforward implementation. Additionally, Singh et al. (Singh et al., 2019) leveraged machine learning-based methods like Support Vector Machine (SVM) and Gaussian Process Regression (GPR) to model surface roughness for biocompatible titanium alloys and aerospace-grade alloys processed on wire EDM, respectively. Atul et al. (Raj et al., 2022) utilized Random Forest and SVM methods to model responses while processing titanium alloy using WEDM. These studies collectively underscore the efficacy of nature-inspired optimization algorithms and machine learning-based techniques in improving optimization processes and modeling responses for complex engineering tasks.

In this study, we aim to analyze EDM outputs using Pareto analysis, specifically targeting scenarios where powders are added, and electrode vibration systems are employed. To achieve this, we develop a suite of machine learning methods tailored for classifying EDM outputs. These models utilize datasets collected through both in-situ and ex-situ monitoring techniques, encompassing various parameters such as material removal rate on both workpiece and tool electrodes, ionization time of the kerosene dielectric, pressure applied to the tool electrode, vibration frequencies recorded via Acoustic Emission (AE) technique, powder concentration in the dielectric, and electrode weight measured before and after the process. Our approach involves constructing 5 machine learning models aimed at a set of Multi-Objective Evolutionary (MOE) algorithms (including NSGA-II, NSGA-III, UNSGA-III, and C-TAEA) for regression output of key EDM metrics: Maximizing Material Removal Rate (MRR), Minimizing Electrode Wear Rate (EWR), and Minimizing Surface Roughness (Ra). We concentrate on Powder Mixed-EDM (PMEDM) processes incorporating alternative vibration and electrode pressure. Additionally, we explore how various manufacturing process parameters impact the models' performance. Furthermore, we introduce a data-driven Deep Learning (DL) model classification method, offering greater interpretability compared to conventional machine learning models. This DL model aims to uncover complex relationships within the dataset, providing insights into EDM output classification.

## 2. Methodology

***Figure 1*** showcases the utilization of both ML models and Multi-objective Evolutionary Algorithms (MOEAs). It comprises several components, including raw dataset features, featurization processes, the employed ML models and MOEs, Pareto frontier analysis, and target regression. This section delves into the processes of dataset collection and curation, feature engineering, the selection of ML algorithms, and, finally, the incorporation of MOEAs to guide the decision-making process.

### *2.1 Data Collection*

The data concerning both HAZ geometry and material, as well as surface defects and their types, were sourced from peer-reviewed publications in advanced manufacturing and materials journals, alongside datasets shared by various industries. Significant attention was paid to studies presenting experimental data relevant to these attributes. Additionally, particular focus was placed on the Die sinking EDM–501–(50A) Spark machine, which features two pump tanks for powder input and exit, along with a vibration system applied to both electrodes. Moreover, details regarding the processing parameters and material properties employed in each experiment were compiled. These compiled details are intended to serve as input variables for our ML models. Additionally, the most effective MOEAs are commonly utilized and tailored for our multi-class objectives. These MOEAs are selected to assess the maximum/minimum area, aiming to optimize the performance of PMEDM while minimizing the occurrence of defects.

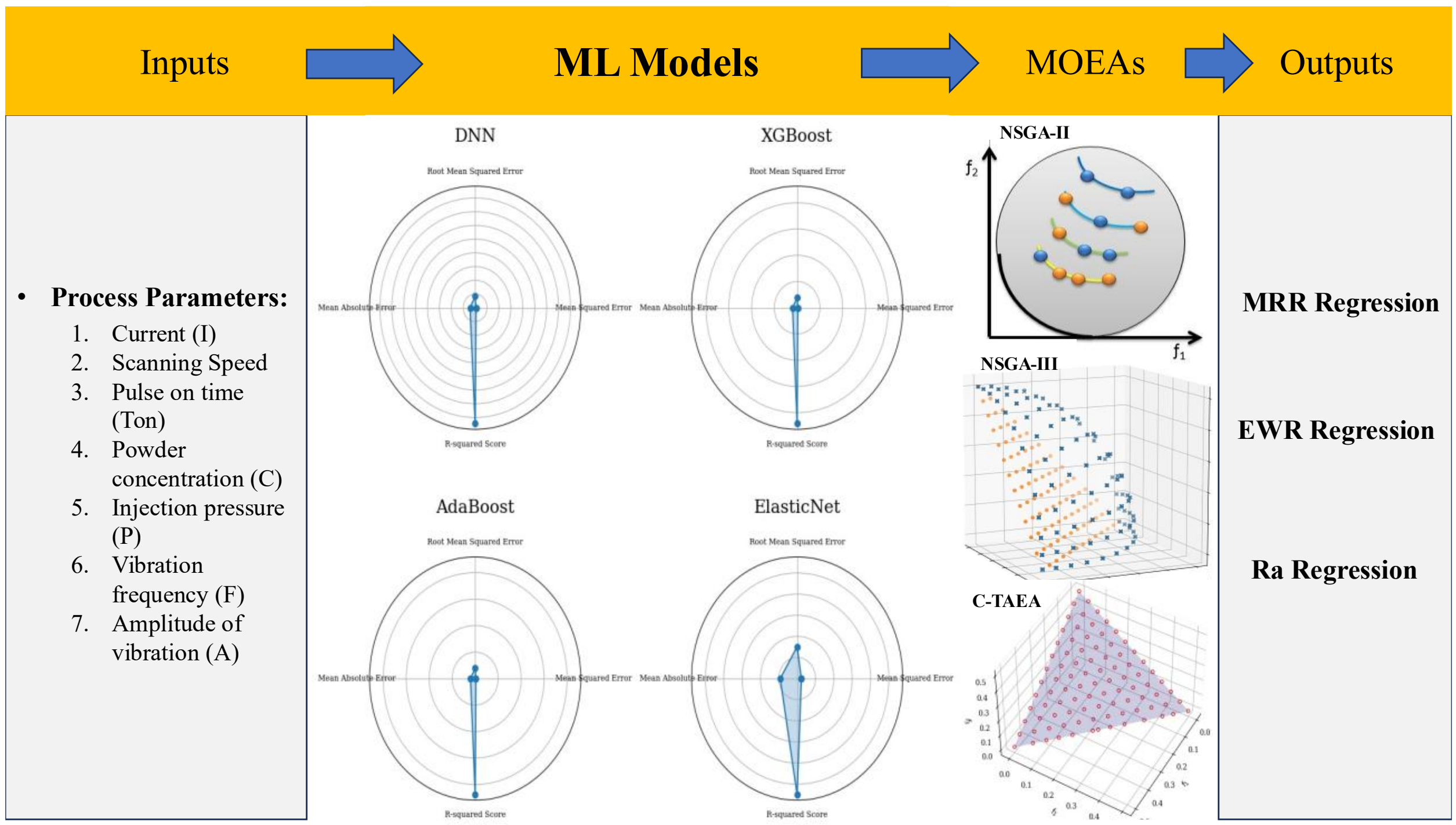

**Figure 1.** Inputs, ML models, and Outputs along with the task implemented in ML+MOEAs combination.

### *2.2 Datasets*

We developed machine learning models using MOEAs based on experimental data collected from tests conducted with the Die sinking EDM–501–(50A) Spark machine. Our dataset consists of approximately 212 data points, each containing information on processing parameters, powder additions, and electrode vibration systems. Considering various parameters affecting HAZ and melt pool characteristics, along with physical properties, our study encompasses EDM process parameters, powder addition systems, and vibration systems. These parameters include Current (I), Pulse on time (Ton), Powder concentration (C), Injection pressure (P), Vibration frequency (F), and Amplitude of vibration (A) at each laser spot.

The comprehensive data collection aims to provide insights into the intricate details of PMEDM and the factors influencing MRR, EWR, and Ra. Subsequently, machine learning algorithms were applied using the aforementioned features to predict various properties of these outputs. We employed different ML regression models to accurately predict MRR, EWR, and Ra through PMEDM with the electrode vibration system. Among the 13 models used were conventional ML models such as 'XGBoost', 'LGBM', 'AdaBoost', 'LogisticRegression', 'DecisionTree', 'RandomForest', 'CatBoost', 'GaussianNB', 'KNN', 'Voting', and 'Bagging'. Additionally, new models including 'DNN', 'SVR with rbf kernel', 'SVR with linear kernel', 'SVR with polynomial kernel', and 'SVR with sigmoid kernel' were incorporated for a comprehensive comparison of ML models in PMEDM outputs regression, a comparison not previously explored in this manner, we selected the four ML models that perfoemd better than the others as DNN, XGBoost, AdaBoost and ElasticNet.

After applying ML regression models, we are integrating MOEAs such as Non-dominated Sorting Genetic Algorithm II (NSGA-II), Non-dominated Sorting Genetic Algorithm III (NSGA-III), Unified Non-dominated Sorting Genetic Algorithm III (UNSGA-III), and onstrained Tournament-based Archive Evolutionary Algorithm (C-TAEA) into our analysis. These MOEAs are established optimization techniques utilized in addressing multi-objective optimization problems. This addition constitutes the second point of novelty in our paper, complementing the comparative ML models. Below, we elucidate the rationale behind the incorporation of these methods as four reasons:

- Handling Multi-objective Optimization
  MOEAs are specifically designed to handle optimization problems with multiple conflicting objectives. In our case, we have multiple output variables (MRR, EWR, and Ra) that we want to optimize simultaneously. MOEAs excel in finding trade-off solutions that balance these objectives efficiently.
- Exploring Pareto Front Solutions
  MOEAs search for solutions along the Pareto front, which represents the set of non-dominated solutions where improving one objective comes at the expense of worsening another. By exploring the Pareto front, we can identify a range of optimal solutions that represent different trade-offs between MRR, EWR, and Ra.
- Enhancing Decision-Making
  The Pareto front solutions provided by MOEAs offer decision-makers a clear understanding of the trade-offs between different performance metrics. This allows stakeholders to make informed decisions based on their preferences and priorities.
- Complementing Machine Learning Models
  While machine learning models excel at prediction tasks, MOEAs offer a complementary approach to optimization. By integrating MOEAs with ML regression models, we can leverage the strengths of both techniques to achieve more robust and effective solutions.

Our aim is to employ ML models for predicting the performance of PMEDM, taking into account the introduction of powders and the utilization of electrode vibration. ***Figure 2*** provides an overview of the correlation between input variables and output metrics, categorized into three key classes typically observed in EDM processes: MRR, EWR, and Ra.

EDM, as an advanced manufacturing technique, harnesses electro-thermal energy to eliminate material through a sequence of evaporation, Marangoni force-induced spattering, and subsequent boiling phenomena due to plasma channel formation. Peak current, a pivotal parameter, plays a critical role in deionizing the dielectric liquid and establishing the plasma channel for energy discharge. Elevated current levels intensify the plasma channel, resulting in increased pressure and energy, thereby exerting a more profound influence on the HAZ and melt pool. Geometrically, higher current levels lead to deeper melt pools, inducing more significant structural alterations in the material. The correlation depicted in ***Figure 2*** underscores that heightened peak current correlates with greater material removal, roughness, and electrode wear rate.

The duration of plasma application initially enhances the outputs, reaching an optimal level, beyond which further increases have a detrimental effect on performance. Furthermore, the concentration of added powders and the gap between machining and discharge contribute to efficiency enhancements in MRR while concurrently reducing EWR and Ra, with correlation coefficients of 0.38 and 0.34, respectively. ***Figure 2*** also underscores the correlation among other EDM parameters, powder concentration, and the vibration system.

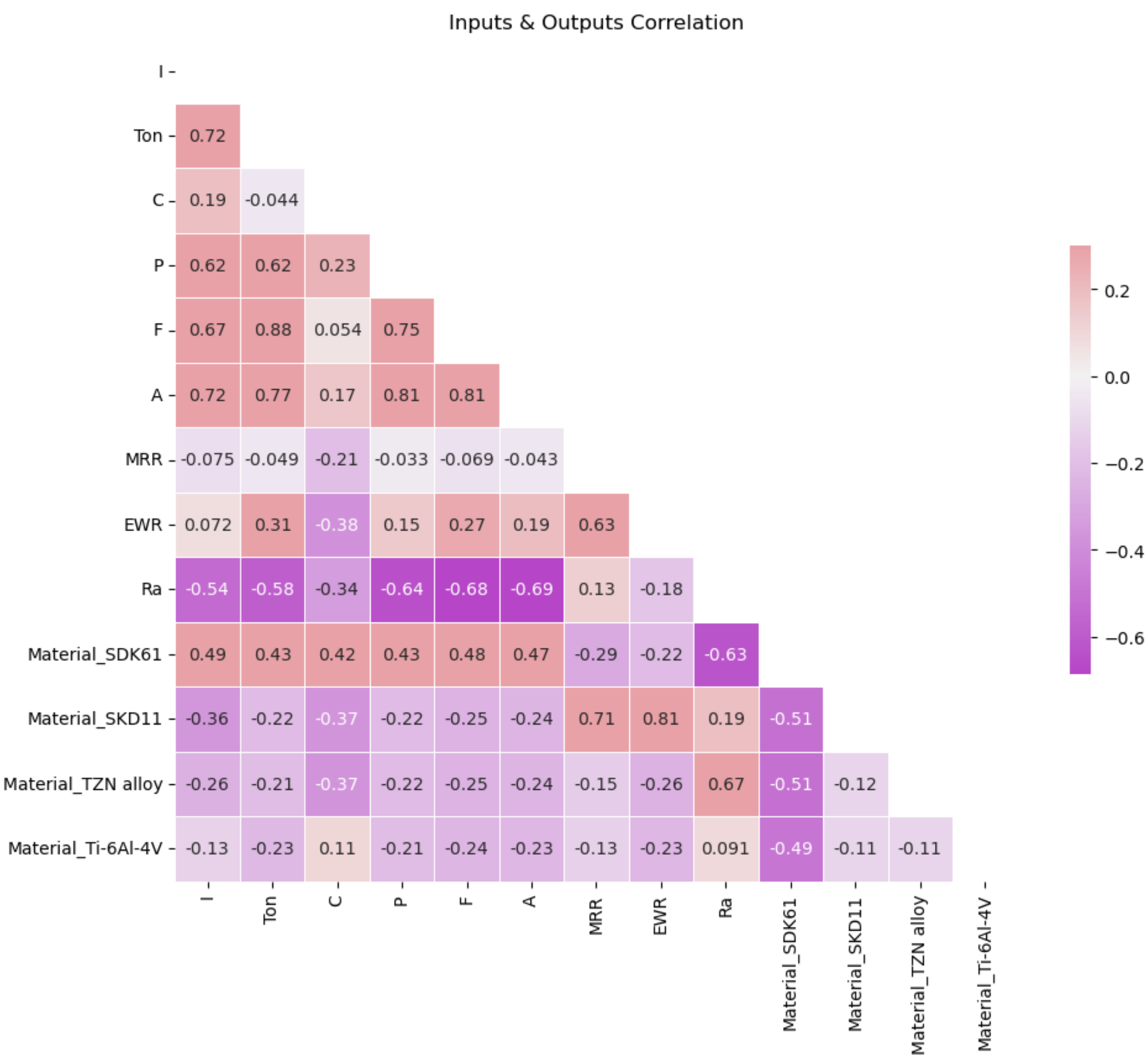

**Figure 2.** Correlation between Inputs and Outputs as MRR, EWR and Ra.

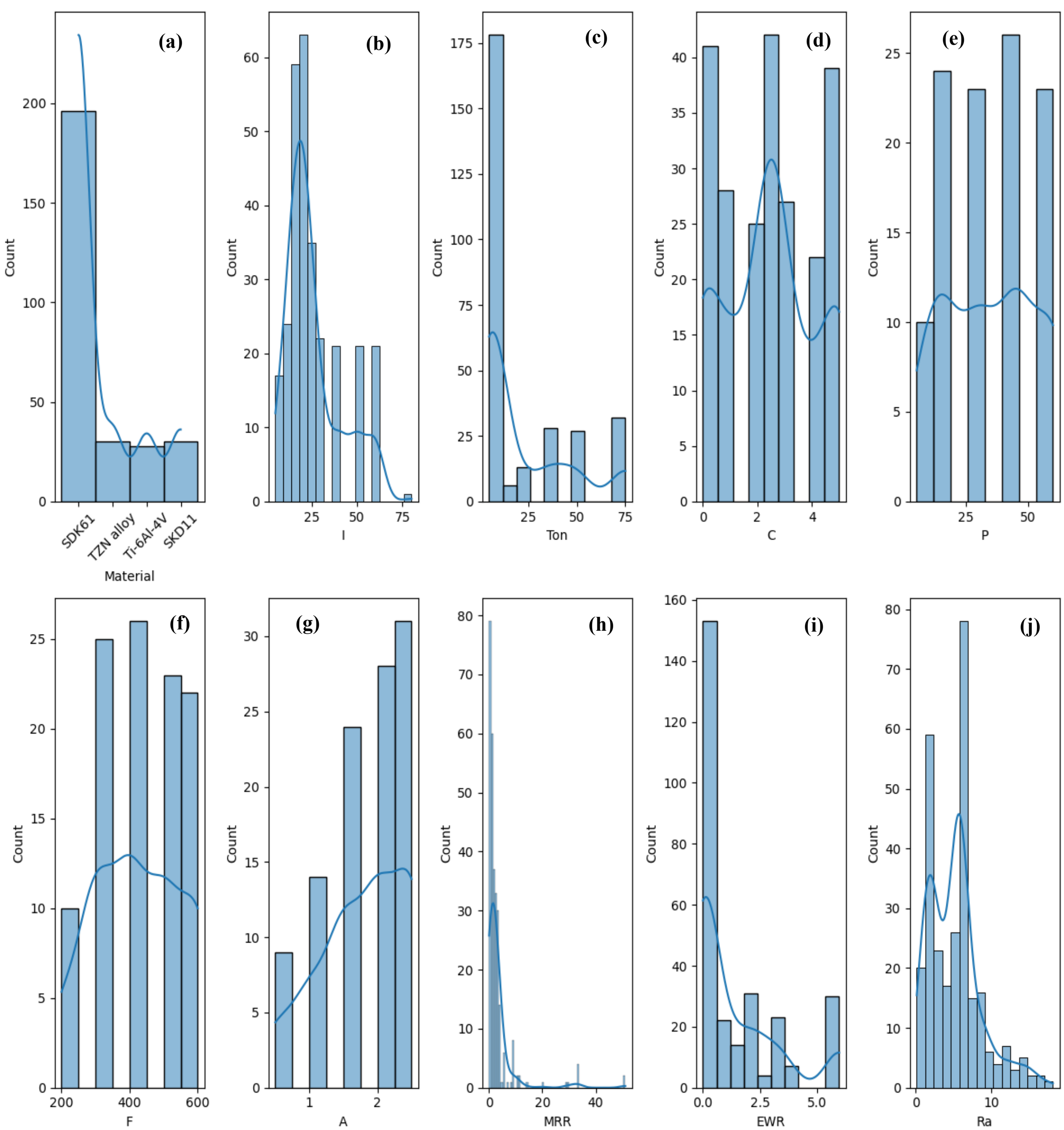

**Figure 3**. Distribution of Feature in EDM while added Powder and Vibration System.

A crucial step in utilizing Machine Learning (ML) models with both numerical and categorical datasets involves cleaning the data as a preprocessing measure. Figure 3 illustrates the distribution of each feature, aiding in the identification of normal or abnormal distributions and pinpointing outliers for effective management and removal if their impact is deemed insufficient.

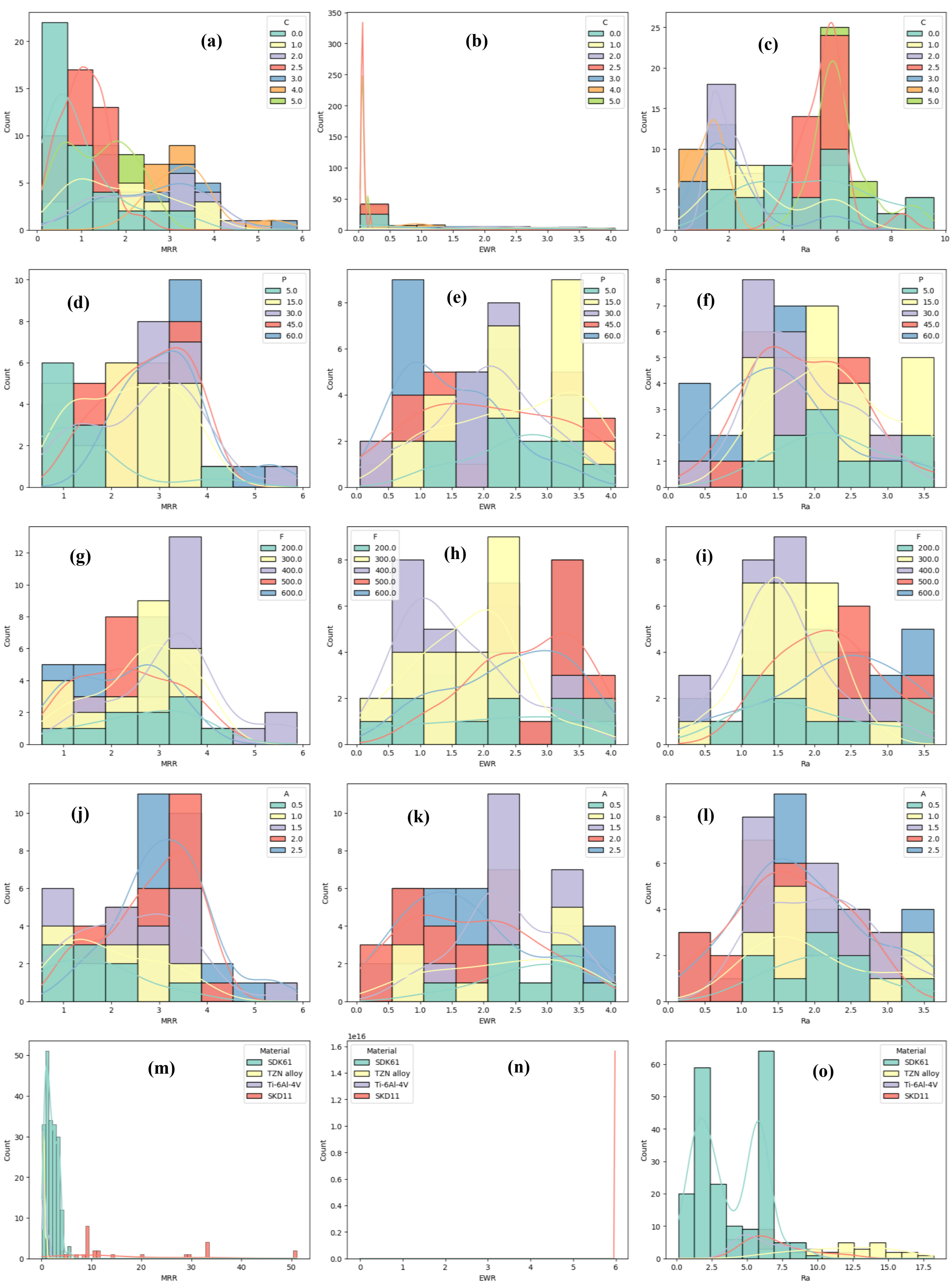

**Figure 4.** Distribution of EDM's Inputs and Outputs by hue of Powders added, Vibration System applied, and Materials considered.

In ***Figure 3*** (a), datasets pertaining to the machining of four material categories—SKD61, SKD11, TZN alloy, and Ti-6Al-4V—are presented. These materials are commonly utilized in the EDM process, with widespread applications in industries such as automotive, aerospace, and biomechanics. The distribution reveals a higher frequency of SKD61 due to its lower cost and broader application, highlighting the challenge of balancing datasets to yield more precise results.

Additionally, two other crucial EDM process parameters—current and pulse on time—are depicted in ***Figure 3*** (b) and (c), reflecting the industry's focus on the production of finished parts and the predominant settings for finishing machining. Considering new trends, the incorporation of powder (Figure 3(d)) and the implementation of a vibration system (***Figure 3*** (e), (f), (g)) are introduced, with their distributions analyzed within a balanced setting to enhance prediction accuracy.

However, the recorded and collected EDM outputs displayed in ***Figure 3*** (h), (i), and (j) capture the intensity of current and electrode wear during the initial stages of the process. These datasets indicate processing for finishing and semi-finishing operations involving rapid material removal within a short timeframe. This suggests that the datasets are tailored for EDM processing with short pulse times, where the plasma channel exerts high pressure and temperature to facilitate evaporation, and the Marangoni force governs the flow out of the melt pool.

This study focuses on employing ML models and optimizing EDM outputs, as illustrated in ***Figure 4***. It examines the impact of adding powders and adjusting parameters of the vibration system to enhance preprocessing and cleaning datasets, thus refining features for ML model input. The addition of powders, ranging in concentration from 0 to 5 g/l (***Figure 4***a, b, c), demonstrates an increase in MRR and a decrease in both EWR and Ra compared to scenarios where no powder is added. This phenomenon aligns with the nature of EDM, where adding powders to the machining gap reduces dielectric coefficient, facilitating easier discharge due to the presence of more conductive materials.

Similarly, applying the electrodes' vibration system (***Figure 4***d-i) showcases varying effects on MRR, EWR, and Ra. Optimal adjustments in injection pressure (P), vibration frequency (F), and amplitude (A) initially increase MRR, but beyond an optimum threshold, their impact diminishes or may even lead to adverse effects (***Figure 4*** d, g, j). Conversely, lower values of these parameters tend to have a more pronounced effect on improving EWR and Ra (***Figure 4*** e, h, k, and ***Figure 4*** f, i, l respectively). Furthermore, common materials used in EDM processes are selected based on their widespread usage across industries and cost considerations. ***Figure 4*** (m, n, o) illustrates that materials with lower hardness and strength are typically removed, encompassing both tool and workpiece electrodes. Materials like Ti-6Al-4V boast higher physical properties and exhibit reduced possibilities of MRR, EWR, and Ra.

### *2.3 Featurization*

In the featurization process, we meticulously craft and compile features from our dataset, which consists of 95 training datasets and 11 testing datasets, to be employed in our ML models for predictive purposes. Given the intricate nature of EDM and the numerous geometric and material properties inherent in HAZ and melt pool phenomena, it is imperative to delineate a substantial number of features to train an ML model for property prediction effectively.

Given that EDM processes involve a blend of numerical and categorical features, incorporating both one-hot encoding and sub-categorical features of PMEDM becomes indispensable. One-hot encoding facilitates the conversion of categorical features, where each value belongs to one of several non-numeric categories, into numeric categories comprehensible by conventional ML models. This encoding transforms a categorical feature with 'n' possibilities into 'n' binary encoding features. In this scheme, for a data sample belonging to a specific category, a '1' is assigned to the corresponding encoding feature, while '0' is assigned to the remaining 'n-1' encoding features. This approach enables us to perform prediction tasks while recognizing the diverse relationships that different heat sources and feedstock material supply methods may have with other features and the prediction target.

### *2.4 Dataset splitting*

In machine learning, it's essential to partition datasets into training and test subsets to evaluate model performance on unseen data. Therefore, we split our dataset into training and testing sets. The models underwent training using the training data, while the test data remained isolated to assess model performance. This partitioning process occurred subsequent to scaling the datasets, as depicted in ***Figure 5***.

Furthermore, we employ k-fold cross-validation to deepen our insight into model performance by integrating multiple training and test partitions. This method involves dividing the dataset into k partitions, where k−1 partitions are utilized for training the model and the remaining partition is held for testing. This process iterates k times, ensuring each partition is used for testing exactly once.

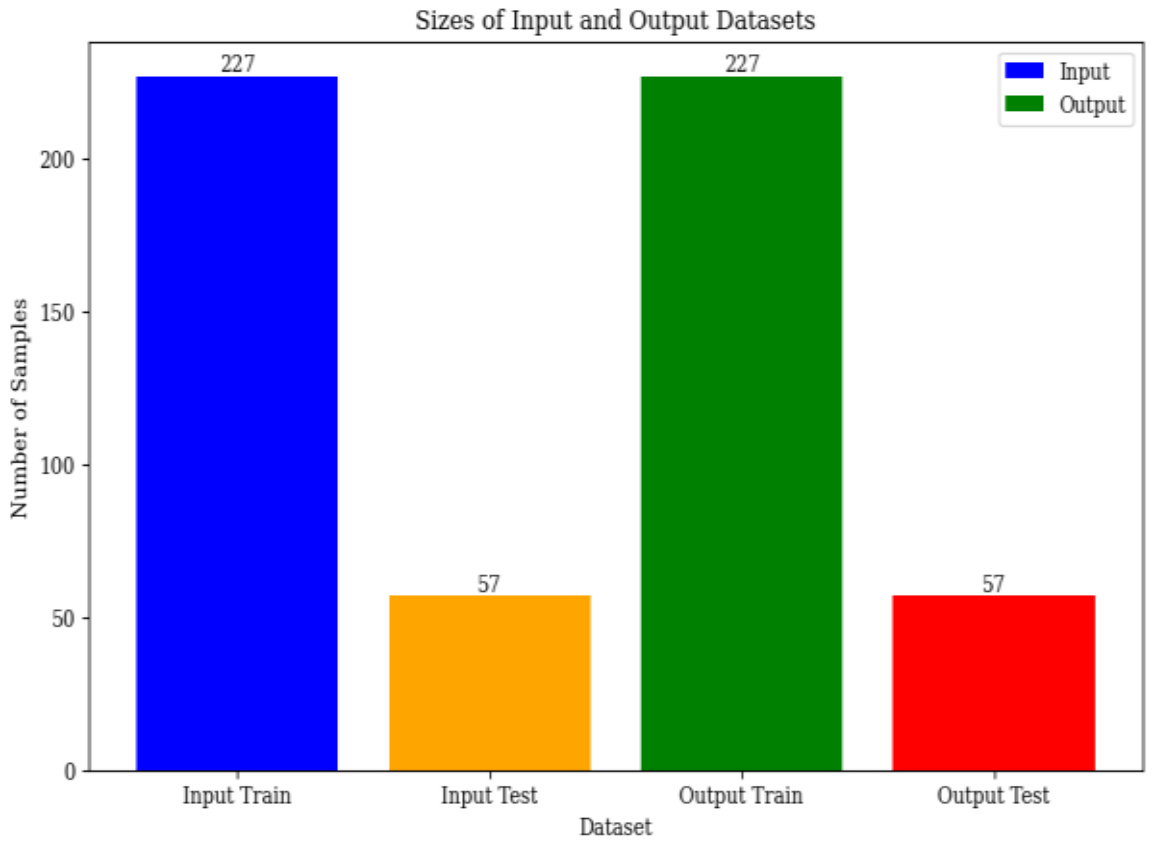

**Figure 5.** Split AM datasets as training and test for both input and output.

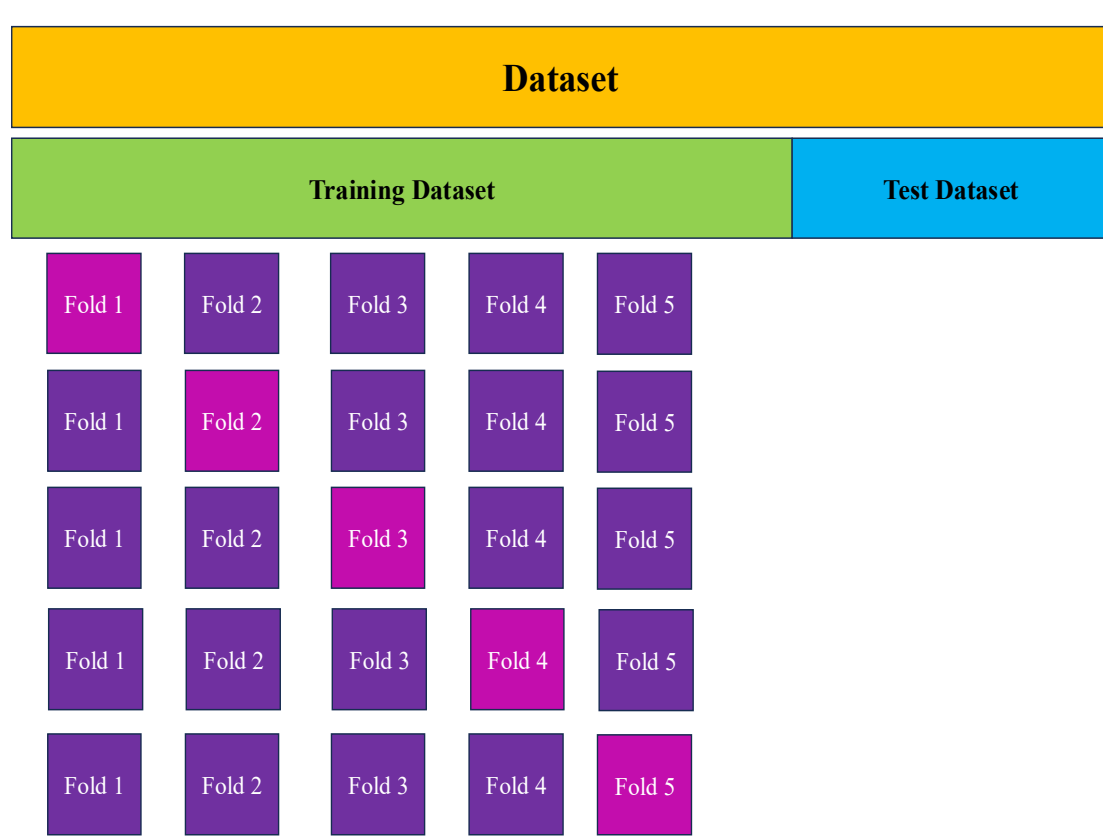

**Figure 6**. $k$-fold cross validation with $k$ =5 was employed on the additive manufacturing (AM) datasets as a preparatory step before applying machine learning models.

The main goal of cross-validation is to prevent overfitting, where a model becomes too closely tailored to the training data, resulting in poor performance on new, unseen data. By assessing the model across multiple validation sets, cross-validation provides a more accurate estimation of the model's ability to generalize to unseen data. However, this method can introduce increased variability in testing models, as it evaluates against individual data points. Outliers within these data points can significantly impact the variability of the testing process. Additionally, k-fold cross-validation can be computationally intensive due to its iteration over the dataset's size. To address these challenges and prevent overfitting, we chose k=5, as it demonstrated highly effective performance with minimal errors and overfitting, as depicted in ***Figure 6***.

### *2.5 ML Models*

Within the domain of multi-physics EDM, where time and cost factors are paramount, there is a growing interest in exploring machine learning (ML) models as promising alternatives to traditional experimental and numerical methods. This study aims to evaluate the effectiveness of various ML models using pre-segmented experimental datasets, which are categorized into linear and non-linear algorithms. In this section, we provide a concise overview of these models, reserving a more comprehensive discussion on effective metric methodologies for subsequent sections. The results and discussions section will then present the outcomes of these models, followed by a comparative analysis in the metrics section to determine their individual efficiencies.

### 2.5.1 Linear Regression Algorithm

A regression algorithm forecasts its output by employing a linear predictor function that merges a set of weights with the feature vector. This function is mathematically represented by the equation:

$$y = f(\vec{w}.\vec{x}) = f(\sum_{j} w_j . x_j)$$

Here, $y$ represents the output, $f$ is the activation function, $\vec{w}$ denotes the weights, $\vec{x}$ is the feature vector, and $\vec{w}.\vec{x}$ represents the dot product of weights and features.

Given the prevalent usage of linear machine learning (ML) algorithms, this study predominantly concentrates on utilizing two well-established models: Support Vector Machine (SVM) with a linear kernel, also recognized as Support Vector Regression (SVR).

#### *2.5.1.1 ElasticNet*

ElasticNet is a linear regression algorithm that combines the penalties of both Lasso and Ridge regression methods. It is particularly useful when dealing with high-dimensional data where many features may be irrelevant or redundant. ElasticNet introduces two parameters, α and λ, where α balances between Lasso (L1) and Ridge (L2) penalties, and λ controls the overall strength of regularization. This algorithm effectively addresses multicollinearity and feature selection issues, making it versatile for various regression tasks.

### 2.5.2 Non-Linear Regression Algorithm

A non-linear regression algorithm serves as a robust machine learning technique designed to capture intricate relationships between input variables and output targets. In contrast to linear regression, which assumes a linear connection between features and outputs, non-linear regression algorithms excel at identifying complex patterns and structures within the data. Our study delves into the exploration of various models geared towards predicting three key outputs – Material Removal Rate (MRR), Electrode Wear Rate (EWR), and Surface Roughness (Ra) – within the realm of Powder Mixed Electrical Discharge Machining (PMEDM). These models encompass tree-based algorithms, Neural Networks (NNs), and Support Vector Regression (SVR), underscoring the versatility and effectiveness of non-linear approaches in regression tasks.

#### *2.5.2.1 Tree-based model*

Tree-based models, which encompass decision trees, random forests, and gradient boosting machines, play a crucial role in machine learning applications for both classification and regression tasks. These models operate by recursively partitioning the feature space into smaller subsets based on feature values, with each split optimizing a chosen criterion such as Gini impurity or information gain. This process results in a tree-like structure where each leaf node represents a final decision or prediction. By combining multiple trees into ensemble models, such as random forests and gradient boosting machines, these approaches enhance predictive accuracy. Overall, tree-based models are valued for their interpretability, flexibility, and ability to capture complex relationships within the data.

##### 2.5.2.1.1 Gradient Boosting Trees (GBT)

Gradient Boosting Trees (GBT) stands out as a robust tree-based algorithm that iteratively enhances model performance by optimizing a loss function. Through sequential addition of decision trees to the ensemble, GBT excels in capturing complex data relationships. In each iteration, a new tree is trained to predict the residuals, bridging the gap between actual and predicted values from previous trees. This iterative refinement process culminates in highly accurate predictions. GBT's strength lies in its adeptness at handling intricate data relationships, making it particularly effective in scenarios with noisy or high-dimensional data. It outperforms other machine learning algorithms in such challenging environments, showcasing its versatility and reliability across various applications.

#### 2.5.2.1.1.1 Extreme Gradient Boosting Machine (XGBM)

XGBoost, or Extreme Gradient Boosting Machine, stands as an advanced implementation of the gradient boosting algorithm, revered for its exceptional speed and performance. Engineered to optimize the gradient boosting process, XGBoost incorporates innovative features such as parallel computing, regularization, and tree pruning. These enhancements elevate the accuracy and efficiency of predictive modeling tasks, making XGBoost a highly sought-after tool in the realm of machine learning.

#### 2.5.2.1.3.2 Adaptive Gradient Boosting Machine (AdaBoost)

AdaBoost, short for Adaptive Boosting, is an ensemble learning method used to build a strong regression model by combining multiple weak regressors. It works by sequentially training a series of weak learners on weighted versions of the training data. In each iteration, the algorithm focuses more on instances that were misclassified in the previous iteration, adjusting its approach to improve performance iteratively.

AdaBoost assigns a weight $\alpha_t$ to each weak learner $h_t(x)$, where $t$ represents the iteration number. The final prediction $H(x)$ is then obtained by summing the weighted predictions of all weak learners:

$$H(x) = sign(\sum_{t=1}^{T} \alpha_t h_t(x))$$

Here, $T$ denotes the total number of weak learners. The sign function ensures that the final prediction is either +1 or -1, depending on the overall weighted sum of the weak learners' predictions.

### *2.5.2.2 Neural Networks (NNs)*

Neural Networks (NNs) represent versatile machine learning algorithms suitable for a wide range of tasks, including regression and classification. These networks consist of interconnected neurons that perform both linear and nonlinear transformations on input data, resulting in refined outputs. The iterative process of backpropagation is employed to adjust weights and biases within the network, aiming to optimize performance by minimizing errors and enhancing accuracy.

### *2.5.2.2.1 Multilayer Perceptrons (MLPs)*

Multilayer Perceptrons (MLPs) constitute a class of neural networks distinguished by their layered architecture, featuring interconnected neurons. Typically, MLPs comprise an input layer, one or more hidden layers, and an output layer. Within the network, each neuron processes input data through weighted connections and applies activation functions to generate output. MLPs find extensive applications in diverse machine learning tasks, such as classification, regression, and pattern recognition, owing to their capability to capture intricate data relationships. The behavior of a node in an MLP can be expressed by the following general equation:

$$a_i = f\left(\sum_{j=1}^{n} w_{ij}.x_j + b_i\right)$$

Where, $a_i$ is the output of the $i$-th node in the layer, $f$ is the activation function applied element-wise, $w_{ij}$ is the weight connecting the $j$-th input to the $i$-th node, $x_j$ is the $j$-th input to the node, $b_i$ is the $i$-th bias term for the node, and $n$ is the number of inputs to the node.

## *2.6 Multi-objective Evolutionary Algorithms (MOEAs)*

Multi-objective Evolutionary Algorithms are optimization techniques used to address problems with multiple conflicting objectives. Unlike traditional single-objective optimization algorithms, MOEAs aim to find a set of solutions that represent trade-offs between the competing objectives, rather than a single optimal solution.

MOEAs operate by maintaining a population of candidate solutions, also known as individuals or chromosomes, and iteratively improving them through the process of evolution. Evolutionary operators such as selection, crossover, and mutation are applied to generate new candidate solutions, which are then evaluated based on their performance with respect to the multiple objectives. The goal is to evolve a diverse set of solutions that collectively cover the Pareto front, representing the trade-off between the conflicting objectives.

### 2.6.1 Multi-objective Evolutionary Algorithms Classification

Multi-objective Evolutionary Algorithms (MOEAs) can be classified into several categories based on their underlying principles, search strategies, and optimization techniques.

#### *2.6.1.1 Pareto-Based Methods*

Pareto-Based Methods are a category of MOEAs that focus on maintaining a diverse set of non-dominated solutions covering the Pareto front. These methods leverage Pareto dominance or non-dominance sorting to evaluate and rank solutions, ensuring that each solution in the set is not inferior to any other solution in all objectives simultaneously. The primary goal of Pareto-Based Methods is to provide a comprehensive representation of the trade-offs between conflicting objectives, enabling decision-makers to explore the solution space and select preferred solutions based on their preferences or requirements.

##### *2.6.1.1.1 Non-dominated Sorting Genetic Algorithm II (NSGA-II)*

NSGA-II, or Non-dominated Sorting Genetic Algorithm II, stands as a prominent player in the realm of multi-objective optimization. This algorithm orchestrates the evolution of a population of candidate solutions, iteratively refining them to enhance their quality across multiple conflicting objectives. Unlike its predecessor, NSGA-II integrates sophisticated selection and reproduction mechanisms, incorporating non-dominated sorting and crowding distance calculations to steer the evolutionary process effectively. Developed as an upgrade to the original NSGA algorithm, NSGA-II addresses previous limitations while introducing novel features aimed at boosting performance. Its key attributes include non-dominated sorting, crowding distance calculation, and elitism, all of which synergize to enable efficient exploration of the solution space, culminating in the identification of a diverse array of high-quality solutions.

##### *2.6.1.1.2 Non-dominated Sorting Genetic Algorithm III (NSGA-III)*

NSGA-III, short for Non-dominated Sorting Genetic Algorithm III, is an evolutionary algorithm designed for solving multi-objective optimization problems. It is an extension of NSGA-II, aiming to enhance its performance and scalability for complex problem domains. NSGA-III utilizes the concept of non-dominated sorting to rank solutions based on Pareto dominance, ensuring that the generated solution set covers the Pareto front efficiently. One of the key enhancements introduced by NSGA-III is the incorporation of reference points to guide the search process. These reference points help in diversifying the solution set and promoting convergence towards the true Pareto front. NSGA-III also employs a mating selection strategy that encourages the exploration of promising regions of the search space while maintaining diversity among solutions.

#### 2.6.1.1.3 Non-dominated Sorting Genetic Algorithm III (NSGA-III)

UNSGA-III, which stands for Unified Non-dominated Sorting Genetic Algorithm III, is a multi-objective evolutionary algorithm designed for solving optimization problems with multiple conflicting objectives. It is an extension of NSGA-III, aiming to address some of its limitations and improve its performance. One of the key features of UNSGA-III is its unified approach to non-dominated sorting. Unlike traditional NSGA variants that use multiple sorting schemes based on different objectives, UNSGA-III employs a single sorting mechanism that considers all objectives simultaneously. This unified sorting strategy helps in reducing computational complexity and improving convergence efficiency. Additionally, UNSGA-III incorporates innovative strategies for diversity preservation and solution selection, ensuring that the algorithm maintains a well-distributed and diverse set of solutions throughout the optimization process. By balancing exploration and exploitation effectively, UNSGA-III is capable of generating high-quality Pareto-optimal solutions for complex optimization problems.

#### 2.6.1.1.4 Constrained Tournament-based Archive Evolutionary Algorithm (C-TAEA)

C-TAEA, or Constrained Tournament-based Archive Evolutionary Algorithm, is a multi-objective evolutionary algorithm (MOEA) designed to solve optimization problems with constraints. It builds upon traditional tournament-based evolutionary algorithms, integrating mechanisms to handle constraints efficiently. One of the distinguishing features of C-TAEA is its tournament selection mechanism, which operates by selecting individuals from the population based on their fitness values and constraint violation scores. This selection process ensures that only the fittest individuals, considering both objective values and constraint violations, are chosen for reproduction and survival.

Moreover, C-TAEA employs an archive mechanism to maintain a diverse set of non-dominated solutions throughout the optimization process. This archive stores the best solutions found so far, ensuring that the algorithm explores a wide range of feasible and near-optimal solutions. To handle constraints effectively, C-TAEA utilizes penalty functions or repair mechanisms to guide the search towards feasible regions of the solution space while maintaining diversity and convergence towards the Pareto-optimal front.

### *2.7 Evaluation Metrics*

To evaluate the performance of our machine learning models, including those integrated with Multi-objective Evolutionary Algorithms (MOEAs), on unseen data, we initially randomized our datasets and conducted 5-fold cross-validation. This technique partitions the data into five subsets, utilizing one subset for validation while the remaining four serve as the training set in each iteration. The overall accuracy of our models is then calculated by averaging the accuracies obtained across all five iterations.

Our dataset revolves around a classification task aimed at identifying outputs in EDM, such as EDM-MRR, EWR, and Ra. To assess the predictive accuracy and performance of our models, we employ various evaluation metrics, including Mean Squared Error (MSE), Mean Absolute Error (MAE), and R-squared ($R^2$). These metrics provide valuable insights into the effectiveness of our models in capturing underlying patterns in the data and making accurate predictions, as summarized in ***Table 1***.

**Table 1.** Machine Learning Evaluation Metrics.

| Metrics | Symbol | Equation |
|---|---|---|
| **Mean Square Error** | MSE | $MSE = \frac{1}{n}\sum_{i=1}^{n}(y_i - \hat{y}_i)^2$ |
| **Mean Absolute Error** | MAE | $MAE = \frac{1}{n}\sum_{i=1}^{n}\lvert y_i - \hat{y}_i\rvert$ |
| **R-squared** | $R^2$ | $R^2 = 1 - \frac{\sum_{i=1}^{n}(y_i - \hat{y}_i)^2}{\sum_{i=1}^{n}(y_i - \bar{y}_i)^2}$ |

### *2.7 Visualization*

Visualization techniques are essential tools for plotting the procedural processes of machine learning models, providing a clearer understanding and deeper insights into their prediction results. In our study, we utilize visualization techniques, particularly in the context of MOEAs, to trade-off among multiple conflicting objectives such as MRR, EWR, and Ra. These techniques ease the understanding of the problem landscape, identification of dominant relationships among solutions, and validation of optimization results. Additionally, visualization facilitates effective communication among stakeholders by presenting complex multi-dimensional data intuitively. Interactive exploration of the Pareto front enhances user engagement, allowing for a deeper understanding of the problem and ultimately improving the efficiency and effectiveness of the optimization process. In our paper, we assume the use of two common visualization tools: Pareto Front Visualization, which represents non-dominated solutions and visualizes trade-offs between conflicting objectives using techniques like scatter plots or line plots, and Interactive Visualization, which allows users to explore the Pareto front and individual solutions in detail through techniques such as interactive scatter plots or 3D plots. These tools enable users to zoom in on specific regions of interest, filter solutions based on objective values, and interactively explore trade-offs between objectives.

## 3. Results and discussion

In this section, we assess the performance of four ML models on datasets that have not received extensive prior study, aiming to determine the most effective outputs for EDM, considering parameters such as MRR, EWR, and Ra, alongside variables like powder composition and electrode vibration. Initially, we undertake a regression task employing the four ML models and evaluate them using commonly adopted metrics: MAE, MSE, RMSE, and R-squared. Subsequently, for a deeper understanding of model performance, we apply MOEAs utilizing four non-dominated and constrained methodologies, namely NSGA-II, NSGA-III, UNSGA-III, and C-TAEA. It's noteworthy that all benchmark results presented in this paper undergo fine-tuning for optimal hyperparameters through manual search, random search, automated hyperparameter tuning, and tuning with artificial neural networks across multiple runs.

### *3.1 Regression task*

The performance of four distinct machine learning models, namely 'DNN', 'XGBoost', 'AdaBoost', and 'ElasticNet', has been comprehensively evaluated across various metrics such as Accuracy, Mean Squared Error, Root Mean Squared Error, and Mean Absolute Error, as illustrated in ***Figure 7*** and ***Figure 8***. These models were scrutinized for their predictive capabilities in estimating MRR, EWR, and Ra outputs within PMEDM, incorporating a vibration system.

The evaluation encompassed a dataset enriched with features derived from the addition of powder and the utilization of an electrode vibration system. The dataset was partitioned into 227 samples for training and 57 samples for testing, employing a 5-fold cross-validation technique. Each sample was characterized by an input shape of 11 dimensions, with 3 output columns representing MRR, EWR, and Ra for PMEDM assessment. Additionally, the datasets were segregated based on material types, including SKD61, SKD11, SKZ alloys and Ti-6Al-4V.

***Figure 7*** presents a comparative analysis of four machine learning models, aiming to provide insights into their performance. The error metrics, including MSE, RMSE, and MAE, are depicted for XGBoost, AdaBoost, DNN, and ElasticNet models. In the observed range, MSE values range from 0.01 to 0.02, RMSE from 0.08 to 0.15, and MAE from 0.04 to 0.09. Higher values of these metrics indicate a decrease in model performance. Notably, the R-squared metric, indicating the accuracy of the regression models, reflects the superior performance of XGBoost, followed by AdaBoost, DNN, and ElasticNet. The adverse decrease in R-squared values signifies a decline in model performance. This comparative evaluation aids in understanding the strengths and weaknesses of each model, guiding further optimization efforts for enhanced predictive capabilities.

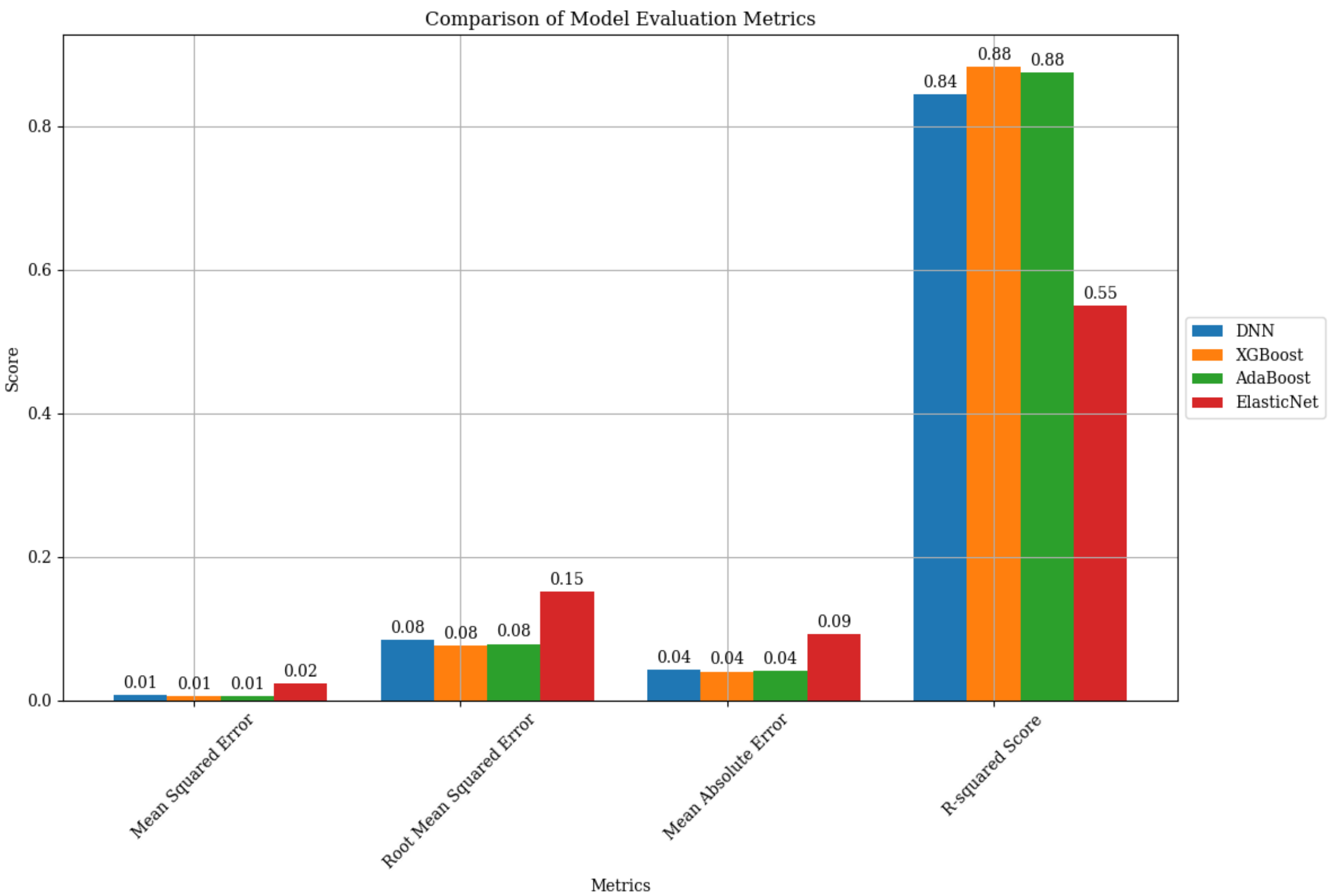

**Figure 7.** Comparison of ML Models ('DNN', 'XGBoost', 'AdaBoost', and 'ElasticNet') Under Regresion Metrics (MSE, RMSE, MAE, and $R^2$).

Additionally, ***Figure 8*** illustrates the utilization of a radar chart for visually assessing the performance of machine learning models across various metrics. Each axis on the radar chart corresponds to a distinct performance metric, including MSE, RMSE, MAE, and R-squared. This chart enables a rapid comparison of the models' performance across these metrics, where higher values indicate poorer performance for error metrics (MSE, RMSE, MAE) and superior performance for R-squared. By offering a comprehensive overview of each model's strengths and weaknesses, the radar chart facilitates informed decision-making and guides subsequent optimization endeavors.

In ***Figure 8***, the performance of each model, including DNN, XGBoost, AdaBoost, and ElasticNet, is represented by the area and vector values on each side of metrics. The smaller area encompassed by DNN, XGBoost, and AdaBoost indicates higher performance, reaching 88%, compared to the conventional machine learning model ElasticNet, which achieves a performance of 55%. This visualization effectively highlights the relative performance of each model, with larger areas and vectors corresponding to better overall performance.

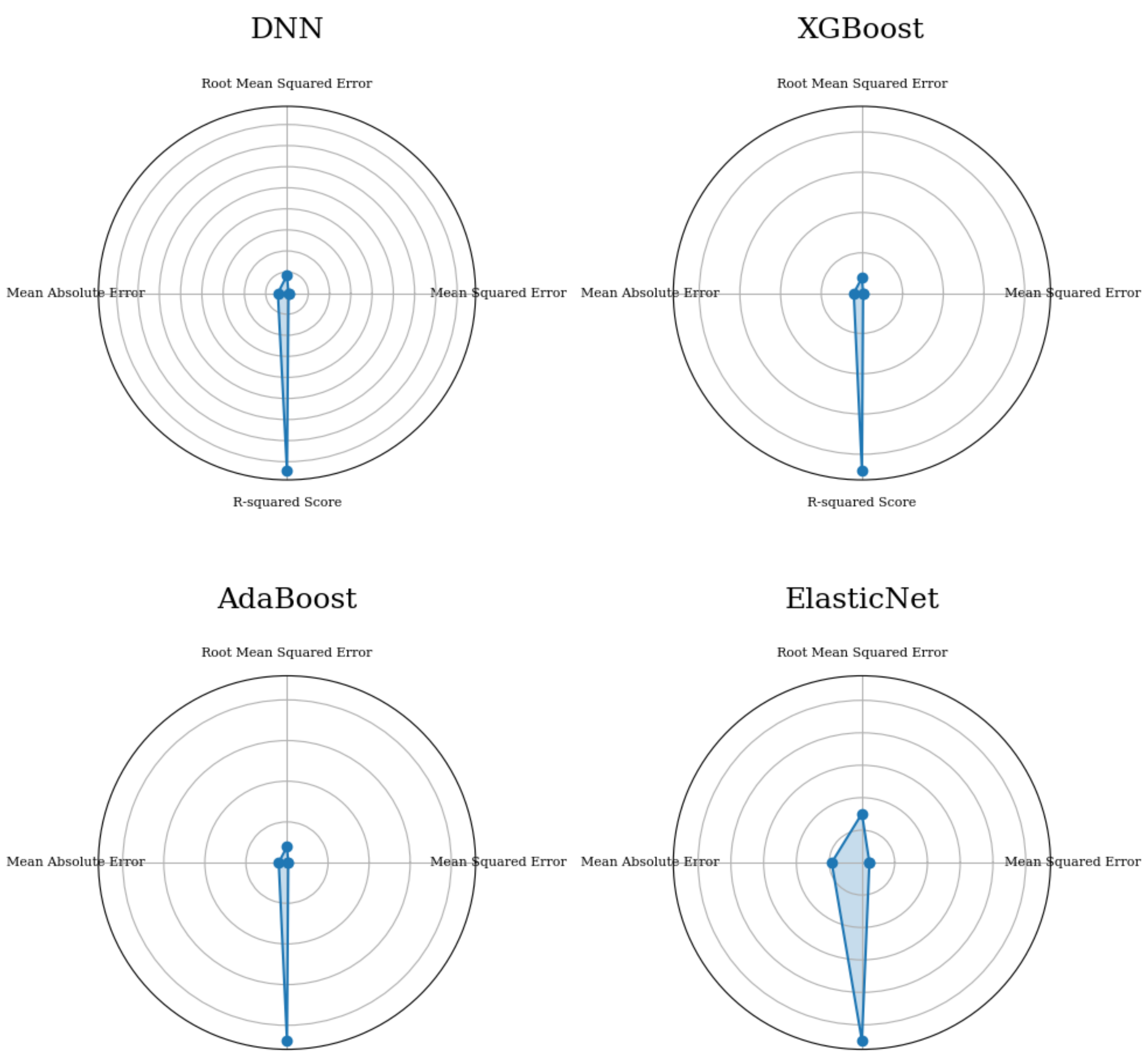

**Figure 8**. Radar Chart illustrating the evaluation of Machine Learning Models under various performance metrics.

## 3.2 Multi-objective Evolutionary Algorithms Task

Multi-objective Evolutionary Algorithms are optimization techniques designed to address problems with multiple conflicting objectives simultaneously. In a typical MOEA task, the problem is formulated with multiple objectives, each potentially having its own constraints and requirements. Solutions to the problem are encoded in a suitable format, and an initial population is generated randomly or using heuristic methods. The fitness of each solution in the population is evaluated based on its performance concerning all objectives, aiming to find a set of non-dominated solutions where no solution is better than another in all objectives. Selection mechanisms, such as tournament selection or roulette wheel selection, are employed to choose solutions for reproduction, while recombination operators like crossover are used to create new offspring. Mutation operators introduce diversity into the population, and replacement strategies maintain diversity and convergence. The algorithm terminates based on predefined stopping criteria, outputting the final set of non-dominated solutions, known as the Pareto front. Post-processing techniques, including visualization and decision-making methods, may be applied to analyze and interpret the obtained solutions further.

In our study, we employed the das-dennis method to generate reference directions, facilitating efficient exploration of the Pareto front by guiding the search towards regions of interest. This method was selected specifically because our problem is designed to optimize three objectives: maximizing MRR (Material Removal Rate), minimizing EWR (Edge Wear Rate), and minimizing Ra (Surface Roughness). By utilizing the das-dennis method, we aim to ensure a well-distributed set of reference directions tailored to our multi-objective optimization task, thereby enhancing the effectiveness of our optimization approach.

### 3.2.1 Non-Dominated Sorting Genetic Algorithm II (NSGA-II)

To inform decision-making based on the prediction values generated by DNN, XGBoost, AdaBoost, and ElasticNet models, two scatter and radar charts were employed to visualize the Pareto front for each ML model under the Non-Dominated Sorting Genetic Algorithm II (NSGA-II). In ***Figure 9***, a detailed depiction of solutions and extended values in scatter and radar plots is provided for the DNN model. Conversely, ***Figure 10*** illustrates that solutions in the Pareto front scatter move in a direct line, with either EWR (f2) or Ra (f3) remaining constant while MRR (f1) increases. Although the number of solutions is lower compared to DNN, the performance of XGBoost under NSGA-II is notable.

Moving to ***Figure 11***, solutions in the Pareto front are concentrated in two distinct regions, with f1 and f2 remaining constant while f3 increases. Despite the reduced number of solutions compared to DNN, AdaBoost demonstrates competitive performance under NSGA-II. Continuing with **Figure *12***, the number of solutions surpasses those of the prior models and is on par with DNN. Notably, the difference between ElasticNet and DNN is approximately linear, with a constant line of Pareto front and reference direction in the increasing direction of MRR. This observation underscores the unique characteristics of ElasticNet compared to the other models evaluated.

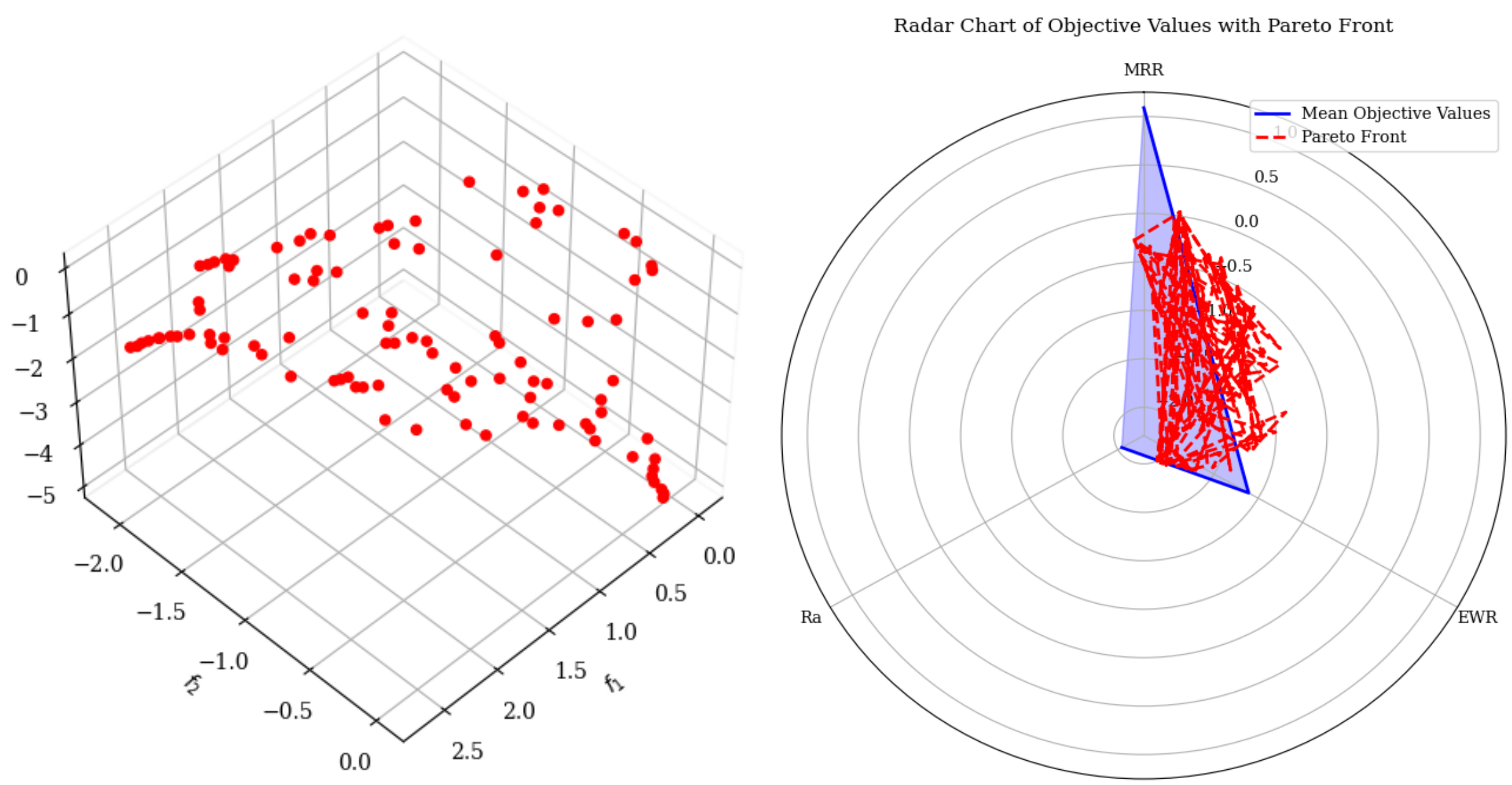

**Figure 9.** NSGA-II Optimization Method on DNN Prediction Results, (a) Pareto Front Solution, and (b) Pareto Front for Reference Direction in Radar Chart.

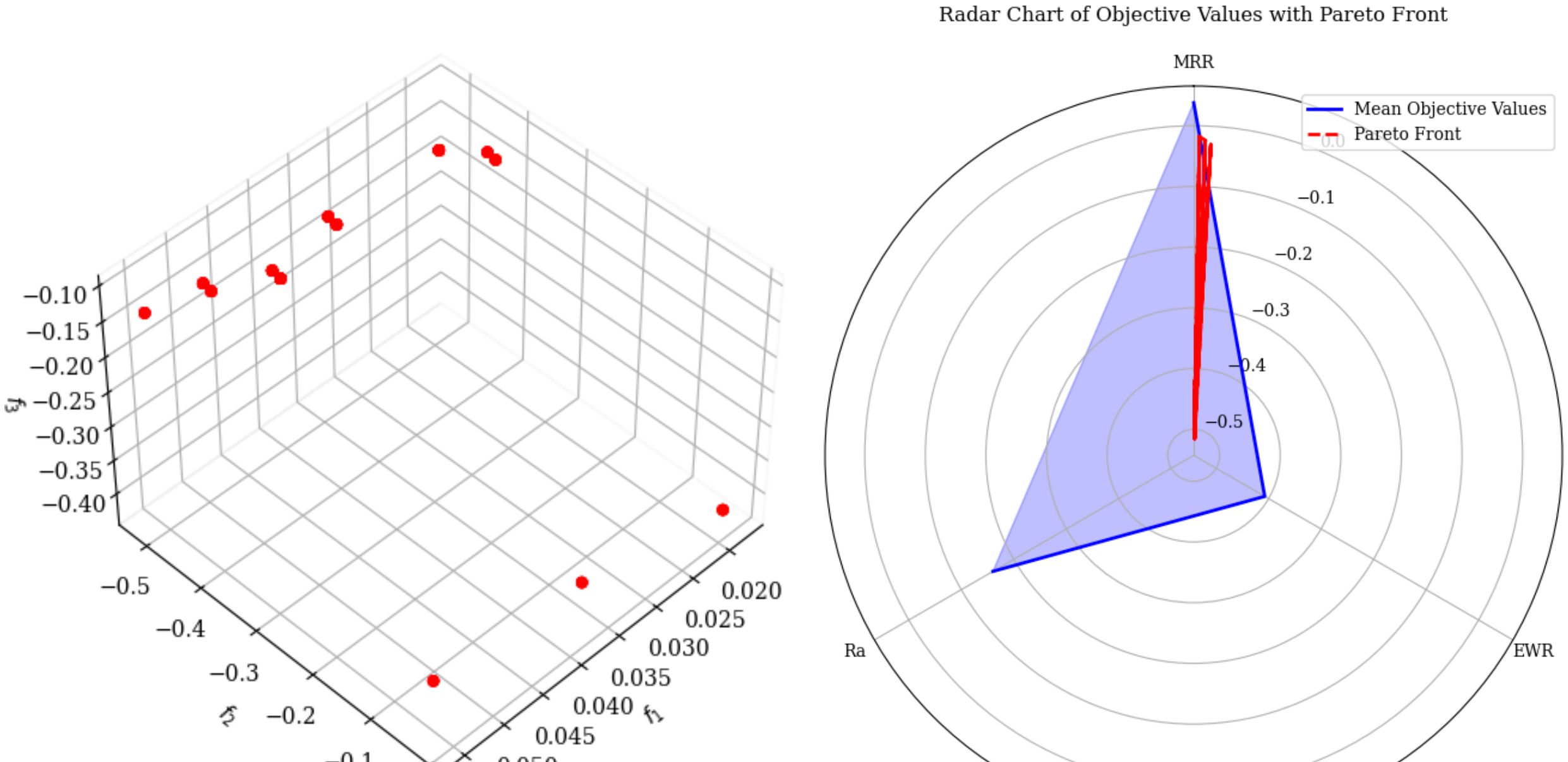

**Figure 10.** NSGA-II Optimization Method on XGBoost Prediction Results, (a) Pareto Front Solution, and (b) Pareto Front for Reference Direction in Radar Chart.

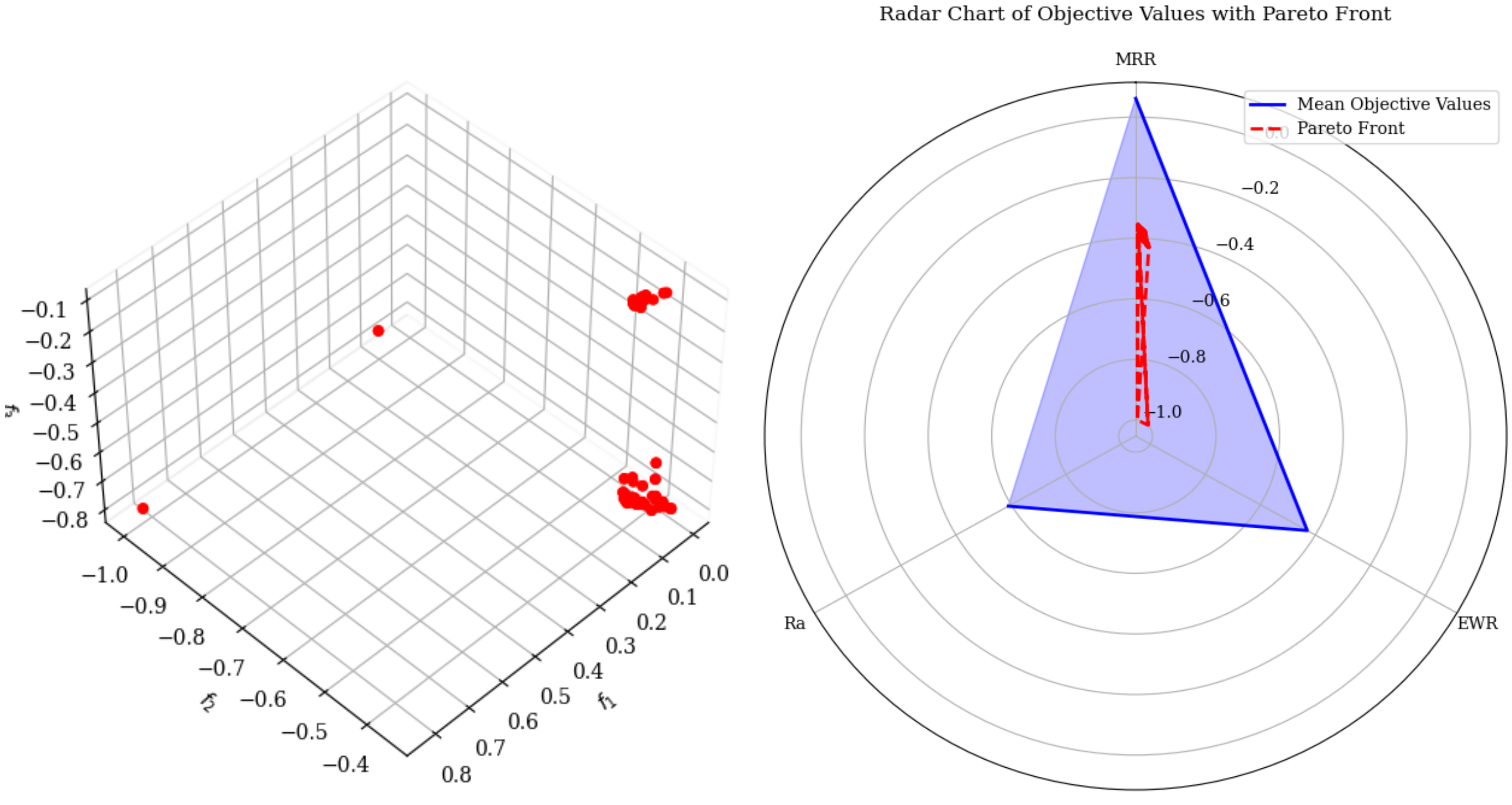

**Figure 11.** NSGA-II Optimization Method on AdaBoost Prediction Results, (a) Pareto Front Solution, and (b) Pareto Front for Reference Direction in Radar Chart.

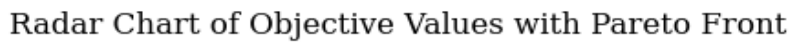

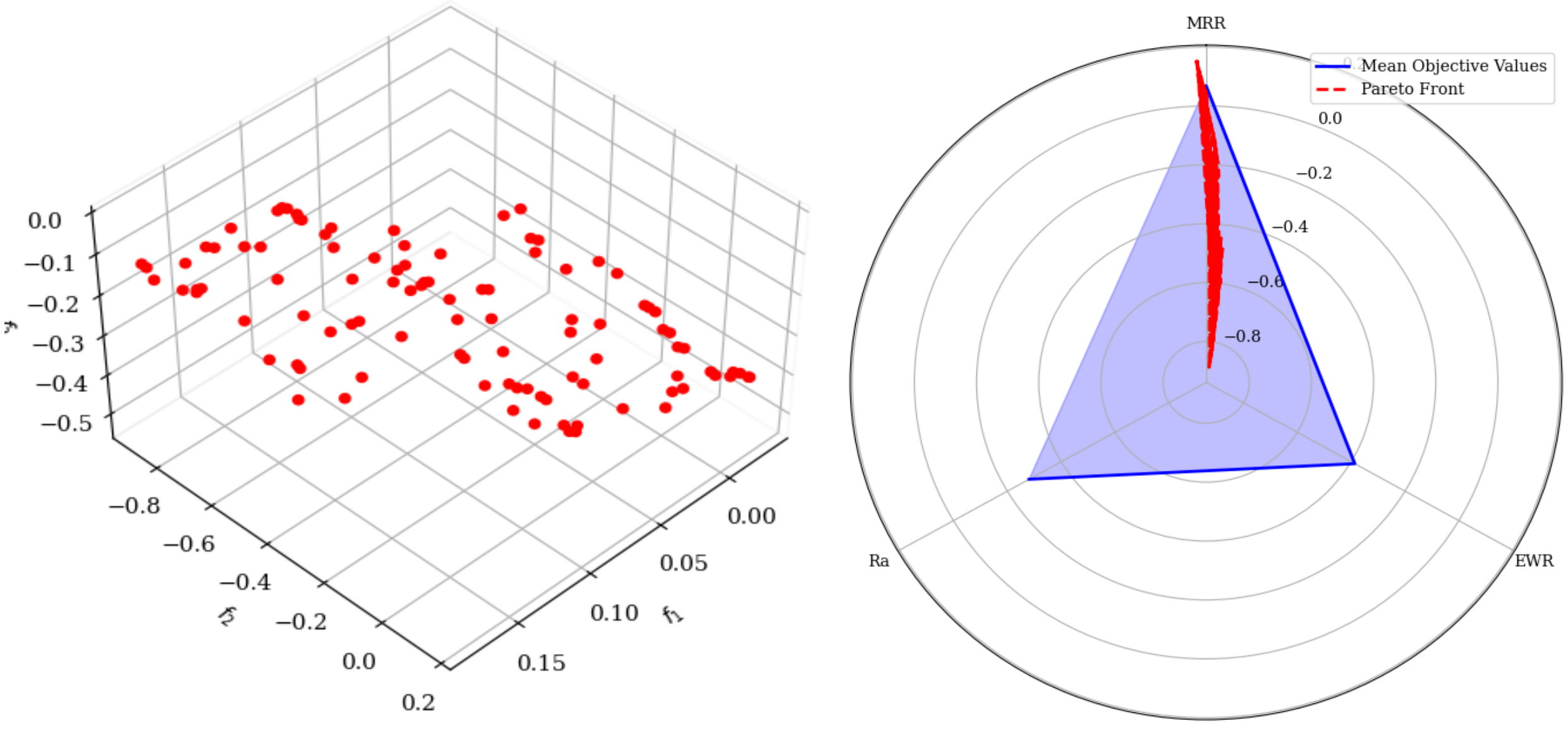

**Figure 12.** NSGA-II Optimization Method on ElasticNet Prediction Results, (a) Pareto Front Solution, and (b) Pareto Front for Reference Direction in Radar Chart.

For reference, the best solutions obtained for each model are as follows:

- DNN: [14.64762913, 0.13596301, 0.23241911]
- XGBoost: [14.64762913, 0.13596301, 0.23241911]
- AdaBoost: [2.27933009, 0.36787944, 0.45151965]
- ElasticNet: [1.19337158, 0.41928543, 0.8496013]

These insights from the Pareto front visualization and best solution analysis provide valuable guidance for selecting the most suitable machine learning model for the given problem domain.

### 3.2.2 Non-dominated Sorting Genetic Algorithm III (NSGA-III)

For Non-dominated Sorting Genetic Algorithm III (NSGA-III), the analysis is similar to that described in the previous content. NSGA-III is an extension of NSGA-II that aims to improve performance by addressing limitations such as maintaining diversity and convergence near the Pareto front.

Incorporating NSGA-III, we observe the following best solutions for each model:

- DNN: [11.6346431, 0.16397513, 0.21736099]
- XGBoost: [1.04965253, 0.59876131, 0.86366217]
- AdaBoost: [2.27933009, 0.36787944, 0.45151965]
- ElasticNet: [1.17311924, 0.49152712, 0.70855333]

Scientifically, these variations in the best solutions can be attributed to the different characteristics and optimization capabilities of each machine learning model. For instance:

- DNN (Deep Neural Network) tends to provide a balanced solution across all objectives, optimizing MRR, EWR, and Ra simultaneously. Its flexibility allows it to adapt to complex patterns and relationships within the data, resulting in competitive solutions (**Figure *13***).
- XGBoost, a gradient boosting algorithm, excels in capturing non-linear relationships and interactions between features. Its best

solution reflects a trade-off between EWR and Ra, possibly indicating a focus on minimizing these metrics at the expense of MRR (**Figure *14***).

- AdaBoost, with its ensemble learning approach, combines multiple weak learners to improve performance. Its solution suggests a trade-off between MRR and Ra, indicating a focus on maximizing MRR while keeping Ra within acceptable bounds (**Figure *15***).
- ElasticNet, a linear regression model combining Lasso and Ridge regularization, offers a compromise between feature selection and model interpretability. Its solution showcases a balance between EWR and Ra, reflecting its ability to handle noisy data and multicollinearity effectively (**Figure *16***).

### 3.2.3 Unified Non-dominated Sorting Genetic Algorithm III (UNSGA-III)

For the DNN model (**Figure *17***), the best solution obtained under UNSGA-III is [3.91009811, 0.41763381, 0.04354874], which differs from the solutions obtained under NSGA-II and NSGA-III. This variation suggests that UNSGA-III might explore different regions of the search space, potentially leading to novel solutions and a more diverse Pareto front.

Similarly, for XGBoost (**Figure *18***), the best solution under UNSGA-III remains [1.04965253, 0.59876131, 0.86366217], consistent with the solution obtained under NSGA-II and NSGA-III. This consistency indicates that the performance of XGBoost might be robust across different MOEAs, highlighting its effectiveness in optimizing the objectives.

The best solutions obtained for AdaBoost (**Figure *19***) and ElasticNet (**Figure *20***) under UNSGA-III, [2.27933009, 0.36787944, 0.45151965] and [1.17967761, 0.48496365, 0.71883236], respectively, show similarities with the solutions obtained under NSGA-II and NSGA-III. This suggests that UNSGA-III may offer comparable performance to these algorithms in terms of exploring and optimizing the Pareto front.

Overall, comparing the performance of different MOEAs, including NSGA-II, NSGA-III, and UNSGA-III, provides valuable insights into their effectiveness in optimizing multi-objective optimization problems. Each algorithm may offer unique advantages and trade-offs, and the choice of the most suitable MOEA depends on factors such as problem complexity, computational resources, and desired optimization outcomes.

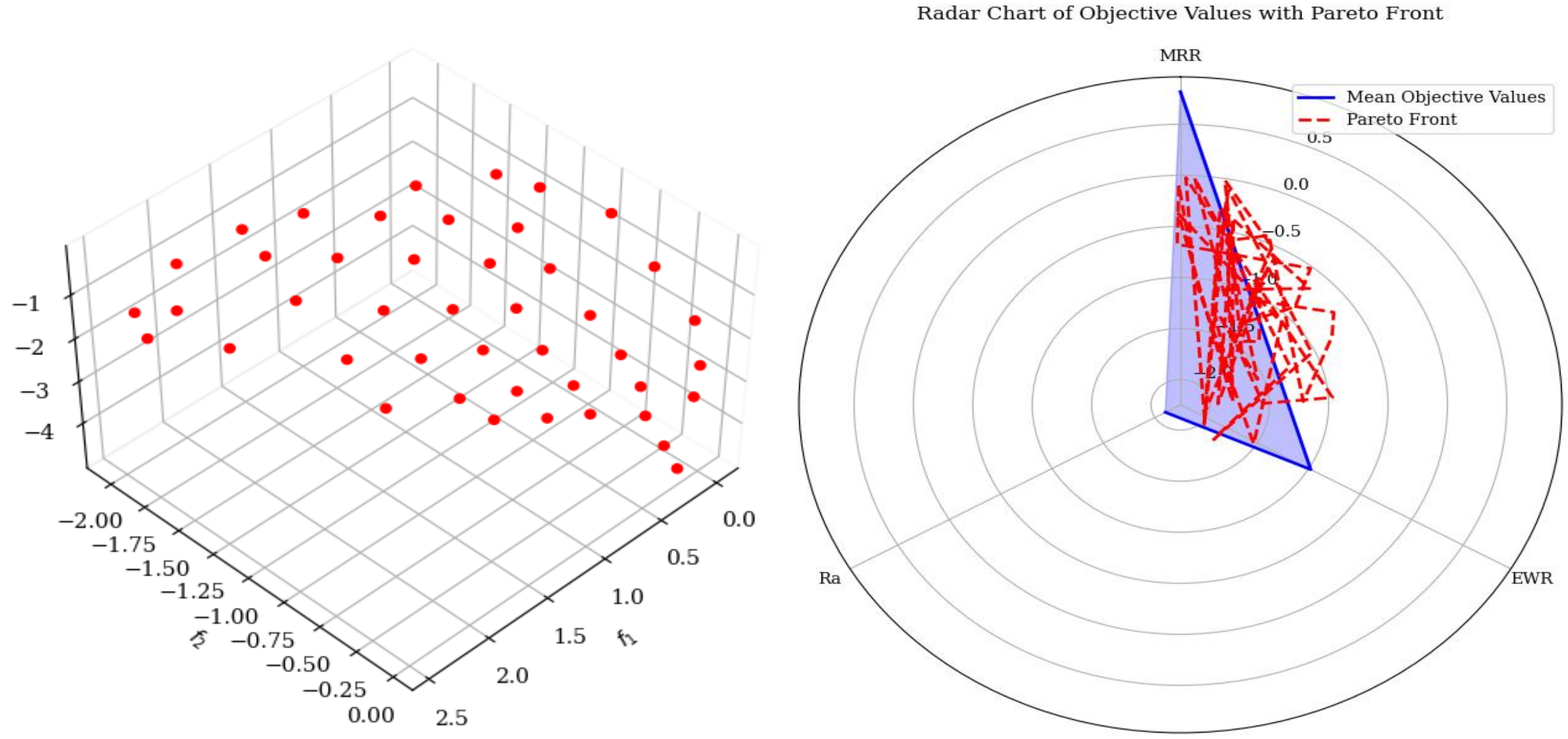

**Figure 13.** NSGA-III Optimization Method on DNN Prediction Results, (a) Pareto Front Solution, and (b) Pareto Front for Reference Direction in Radar Chart.

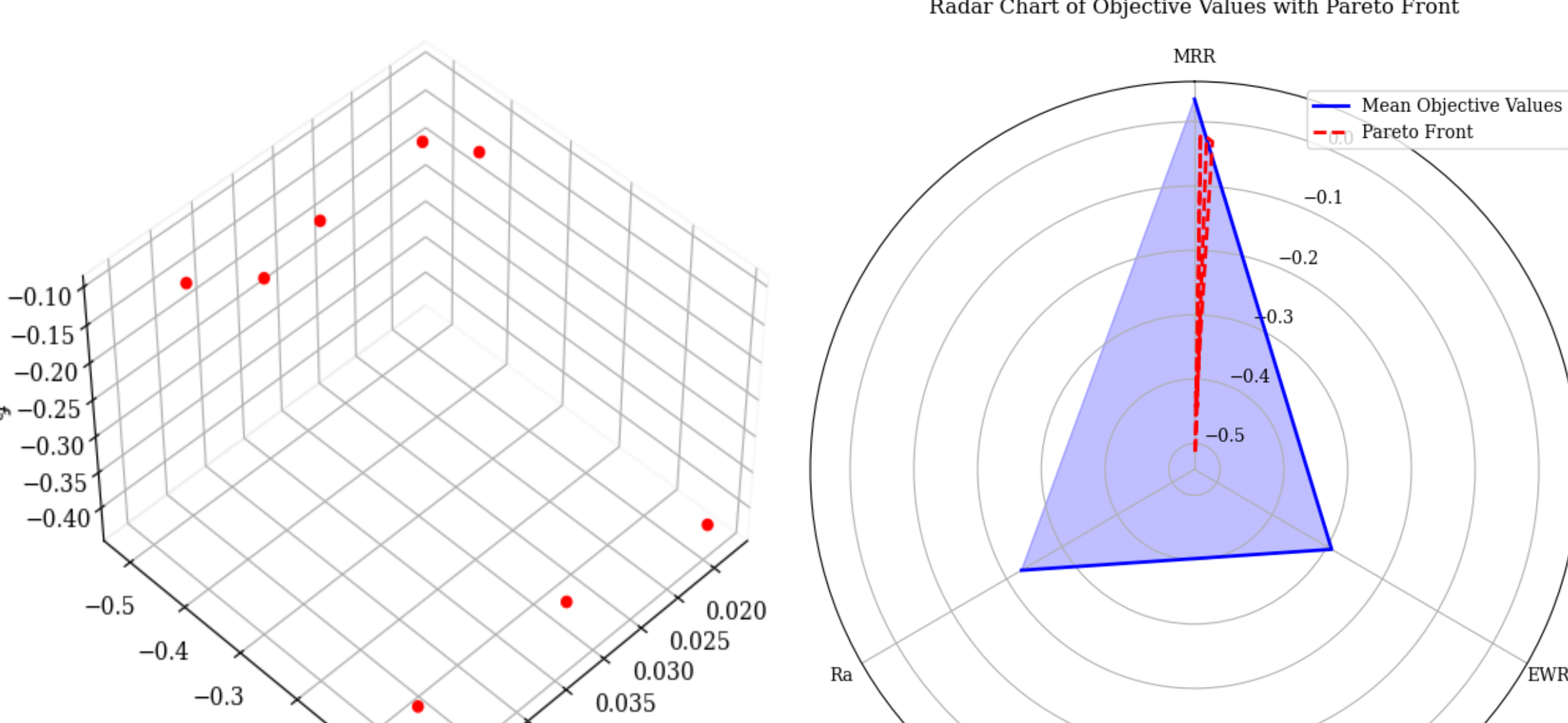

**Figure 14**. NSGA-III Optimization Method on XGBoost Prediction Results, (a) Pareto Front Solution, and (b) Pareto Front for Reference Direction in Radar Chart.

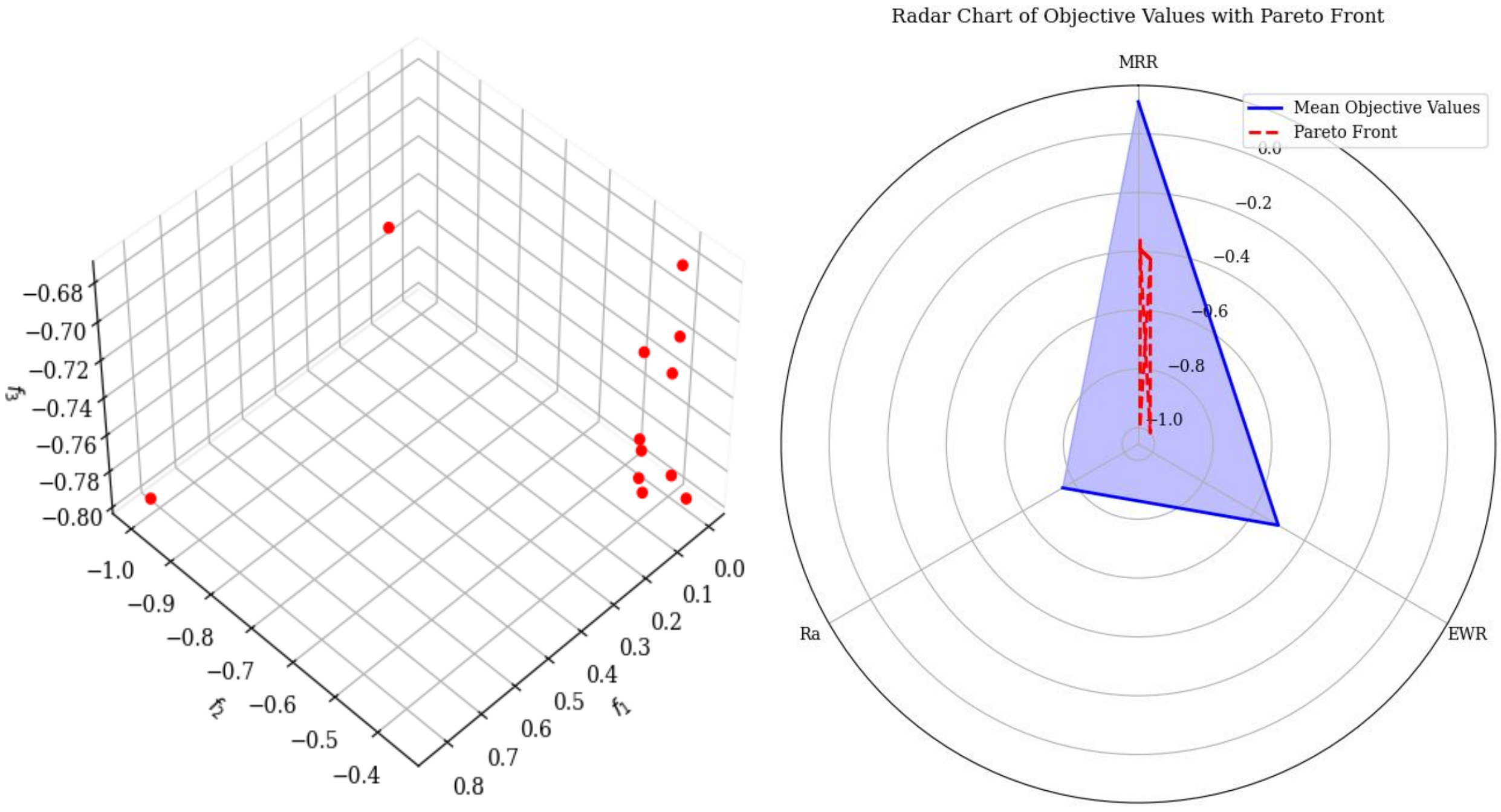

**Figure 15**. NSGA-III Optimization Method on AdaBoost Prediction Results, (a) Pareto Front Solution, and (b) Pareto Front for Reference Direction in Radar Chart.

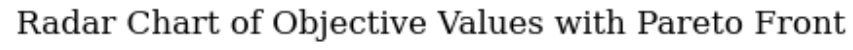

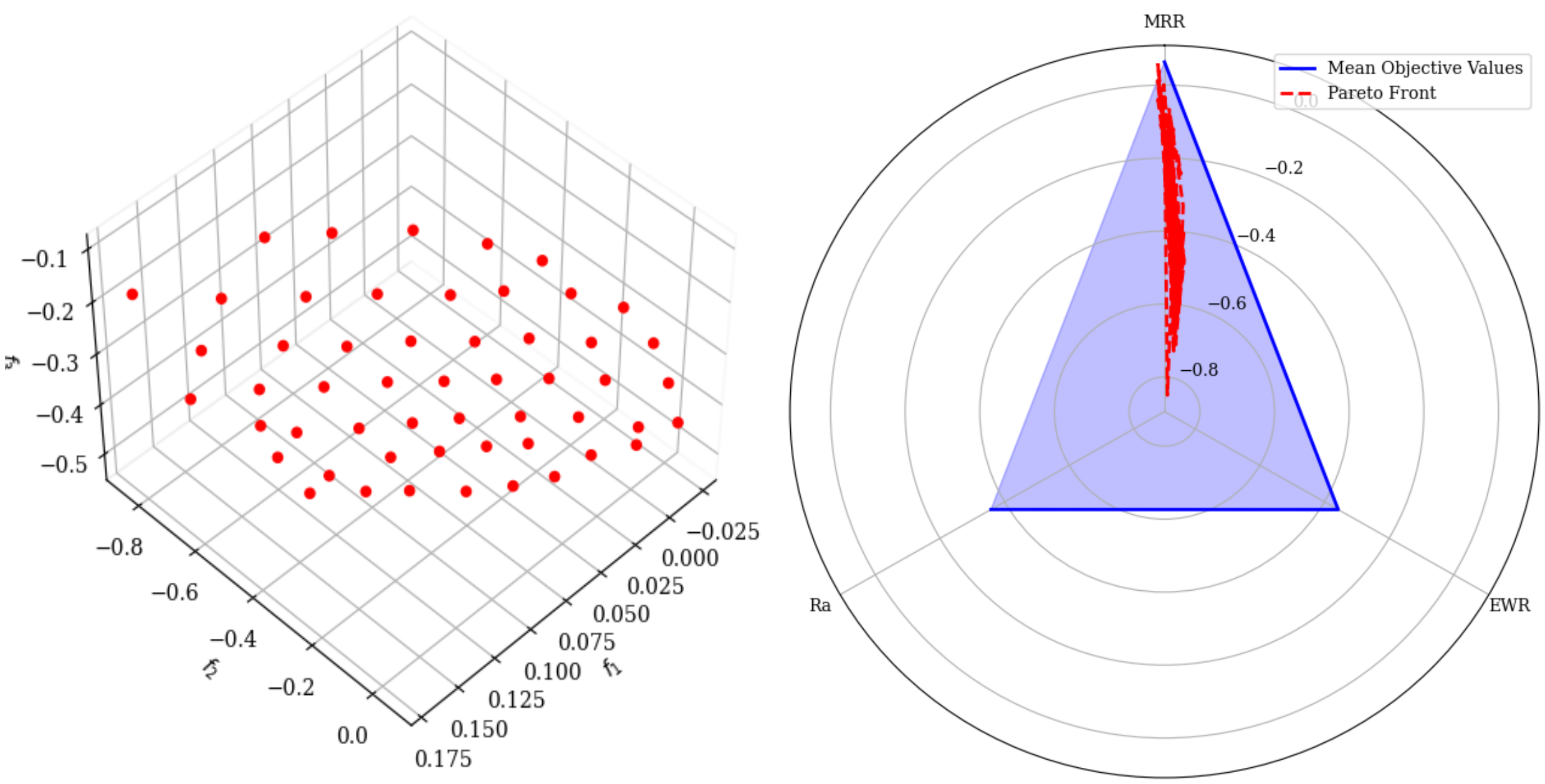

**Figure 16.** NSGA-III Optimization Method on ElasticNet Prediction Results, (a) Pareto Front Solution, and (b) Pareto Front for Reference Direction in Radar Chart.

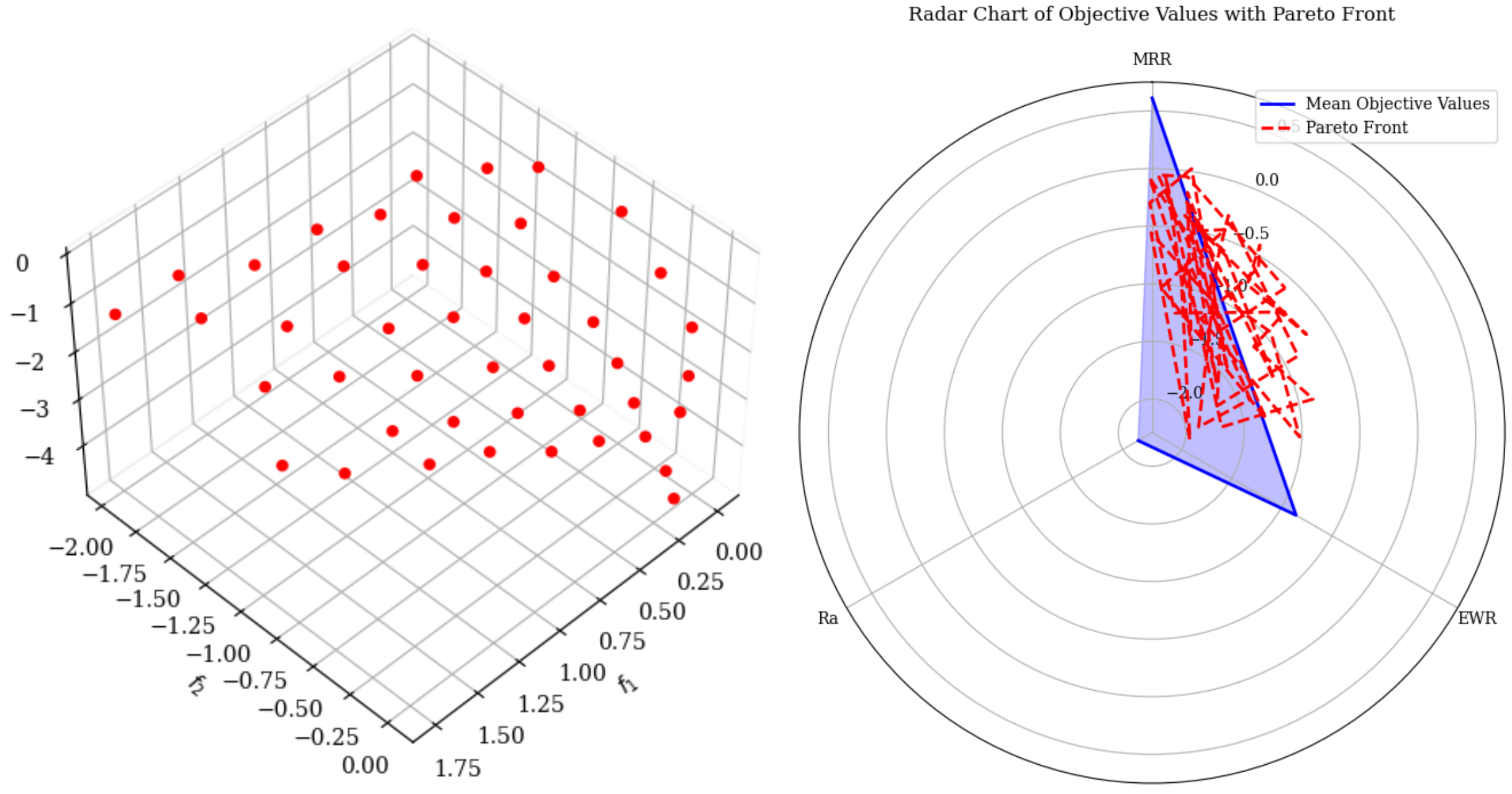

**Figure 17.** UNSGA-III Optimization Method on DNN Prediction Results, (a) Pareto Front Solution, and (b) Pareto Front for Reference Direction in Radar Chart.

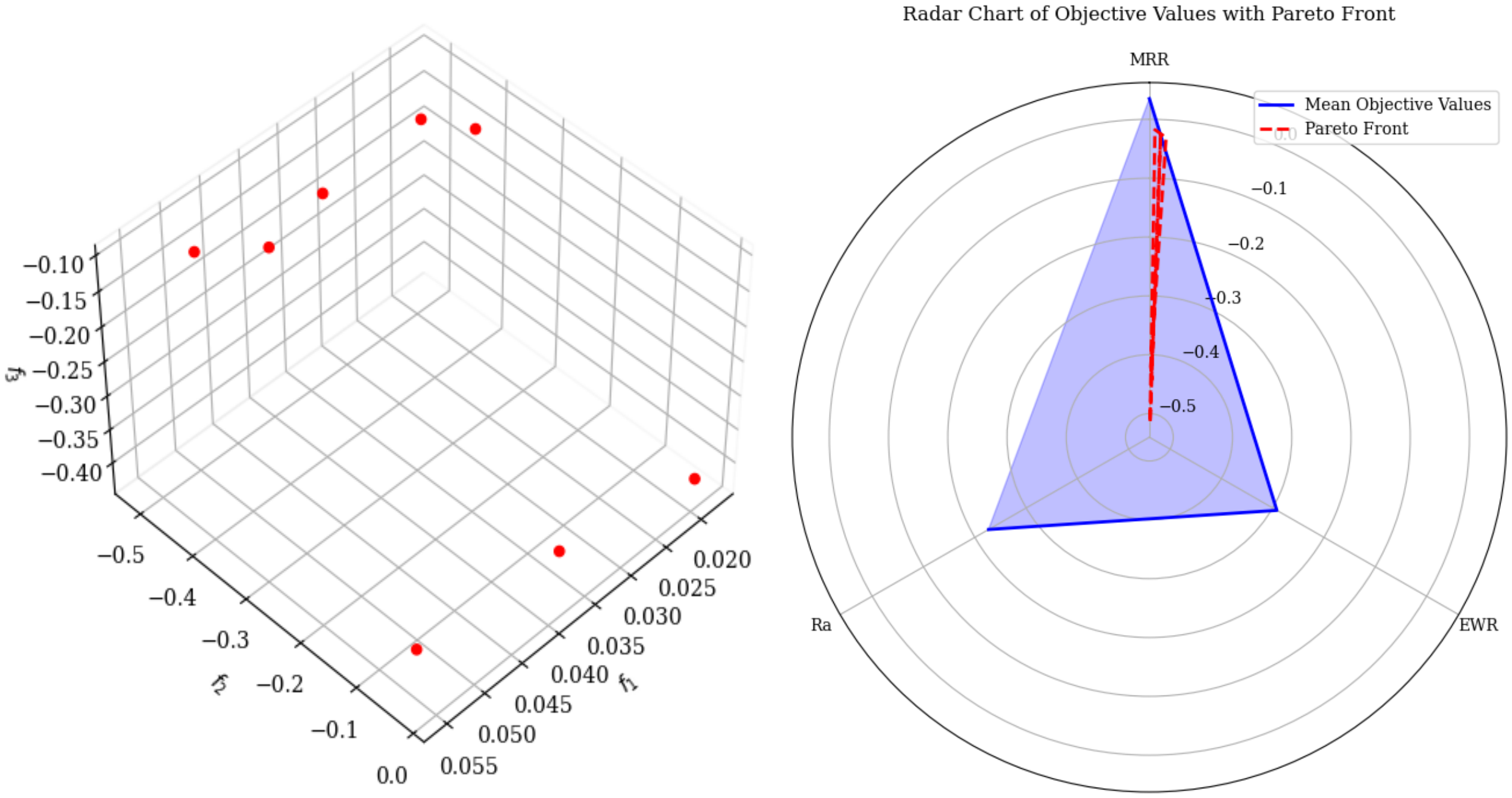

**Figure 18.** UNSGA-III Optimization Method on XGBoost Prediction Results, (a) Pareto Front Solution, and (b) Pareto Front for Reference Direction in Radar Chart.

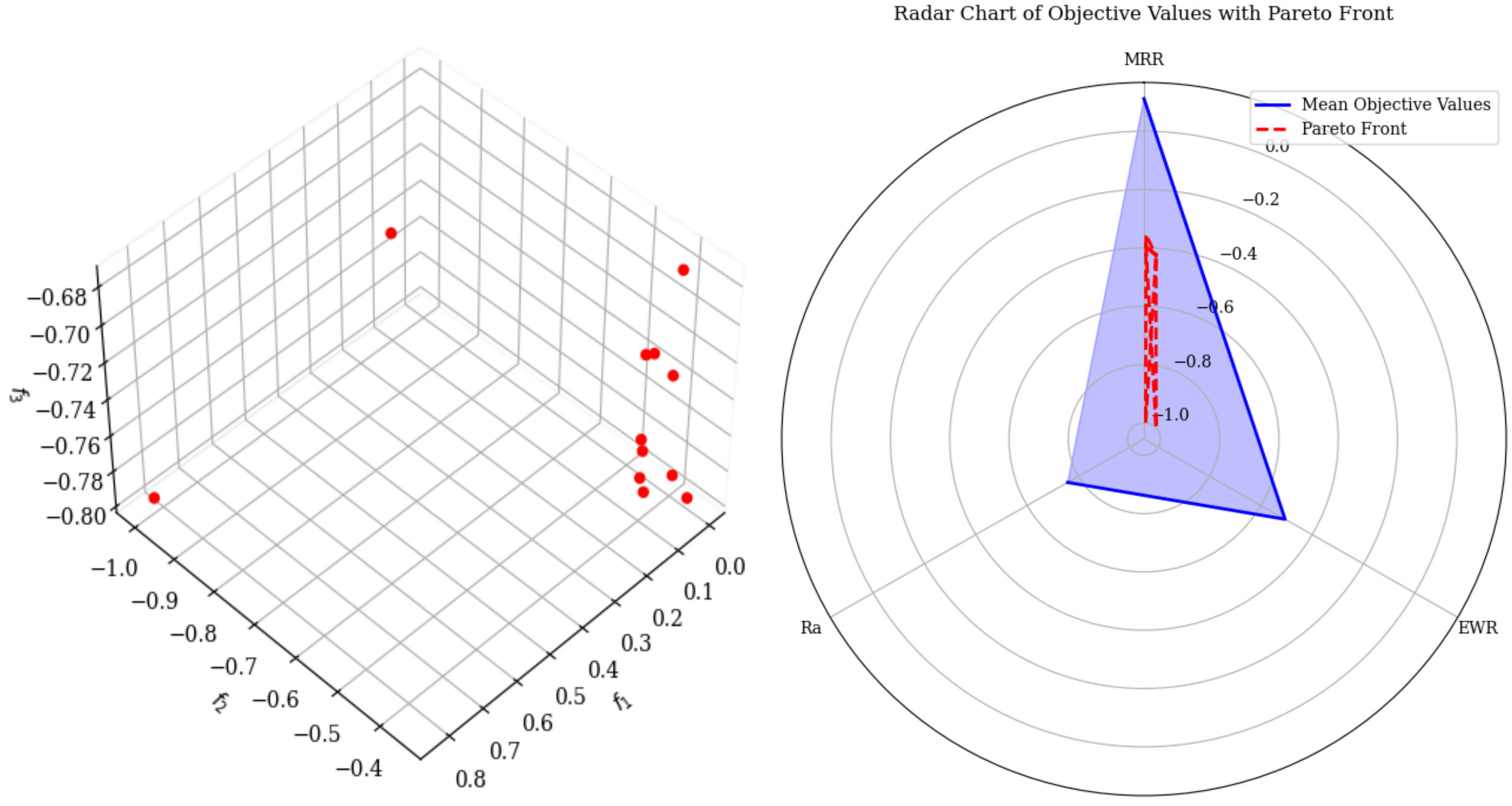

**Figure 19**. UNSGA-III Optimization Method on AdaBoost Prediction Results, (a) Pareto Front Solution, and (b) Pareto Front for Reference Direction in Radar Chart.

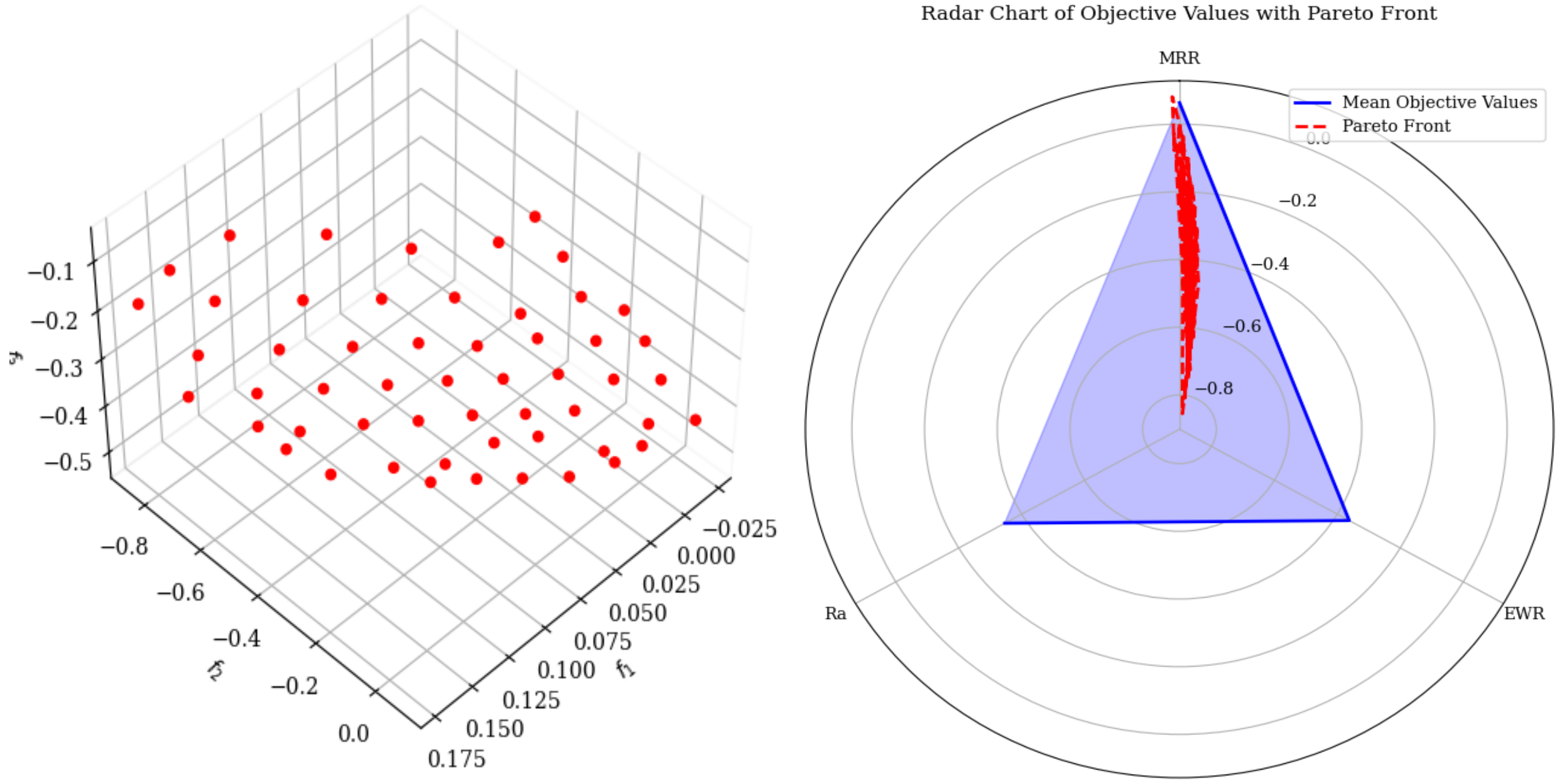

**Figure 20.** UNSGA-III Optimization Method on ElasticNet Prediction Results, (a) Pareto Front Solution, and (b) Pareto Front for Reference Direction in Radar Chart.

### 3.2.4 Constrained Tournament-based Archive Evolutionary Algorithm (C-TAEA)

Comparing the results obtained from the Constrained Tournament-based Archive Evolutionary Algorithm (C-TAEA) with those from other Multi-Objective Evolutionary Algorithms provides valuable insights into their performance and effectiveness in optimizing the problem at hand.

For the DNN model (**Figure** ***21***), the best solution obtained under C-TAEA is [4.29735085, 0.39471319, 0.04880387]. This solution shows slight differences compared to the solutions obtained under other MOEAs, suggesting that C-TAEA might explore different regions of the search space, potentially leading to diverse and novel solutions.

Similarly, for XGBoost (***Figure 22***), the best solution under C-TAEA remains [1.04965253, 0.59876131, 0.86366217]. This indicates that the performance of XGBoost is robust across different MOEAs, highlighting its effectiveness in optimizing the objectives.

The best solutions obtained for AdaBoost (**Figure** ***23***) and ElasticNet (***Figure 24***) under C-TAEA, [2.27933009, 0.36787944, 0.45151965] and [1.17329555, 0.48154035, 0.72506959], respectively, show similarities with the solutions obtained under other MOEAs. This suggests that C-TAEA may offer comparable performance to other algorithms in exploring and optimizing the Pareto front.

Overall, comparing the performance of different MOEAs, including NSGA-II, NSGA-III, UNSGA-III, and C-TAEA, provides insights into their effectiveness in optimizing multi-objective optimization problems. Each algorithm may have unique strengths and weaknesses, and choosing the most suitable MOEA depends on factors such as problem characteristics, computational resources, and optimization objectives.

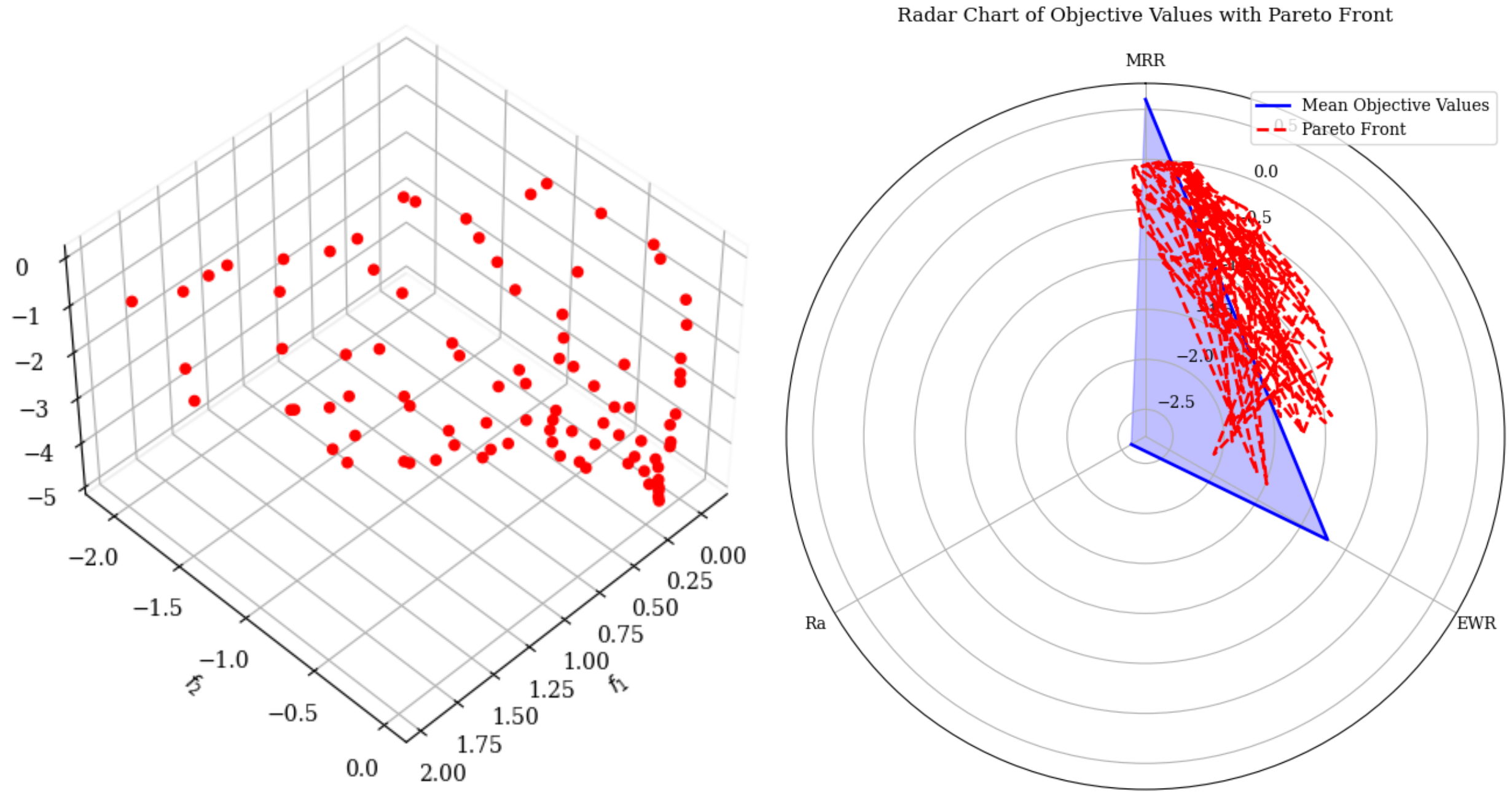

**Figure 21.** C-TAEA Optimization Method on DNN Prediction Results, (a) Pareto Front Solution, and (b) Pareto Front for Reference Direction in Radar Chart.

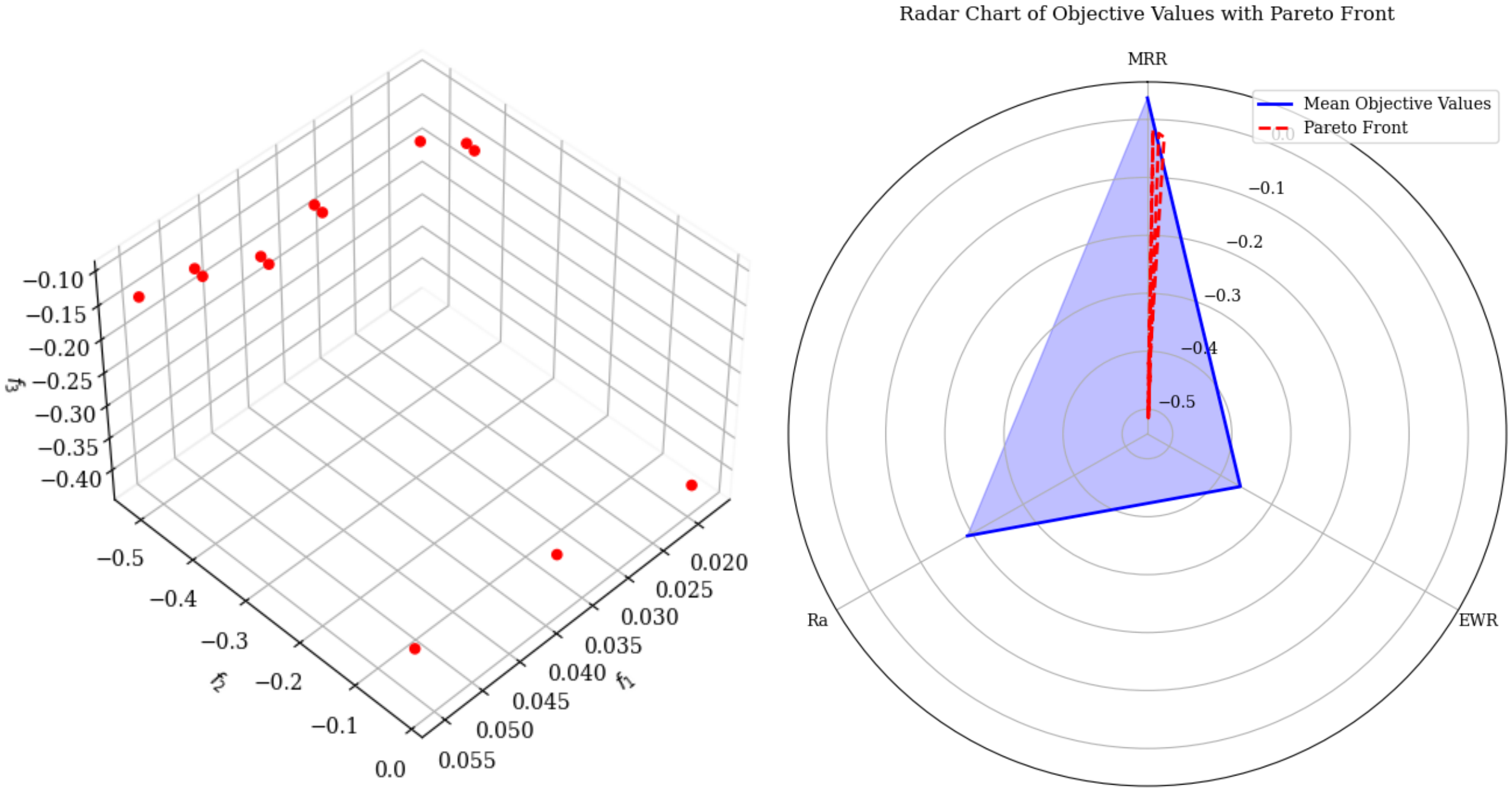

Figure 22. C-TAEA Optimization Method on XGBoost Prediction Results, (a) Pareto Front Solution, and (b) Pareto Front for Reference Direction in Radar Chart.

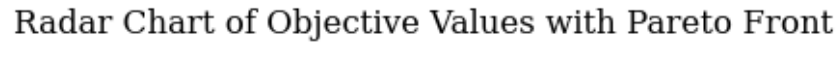

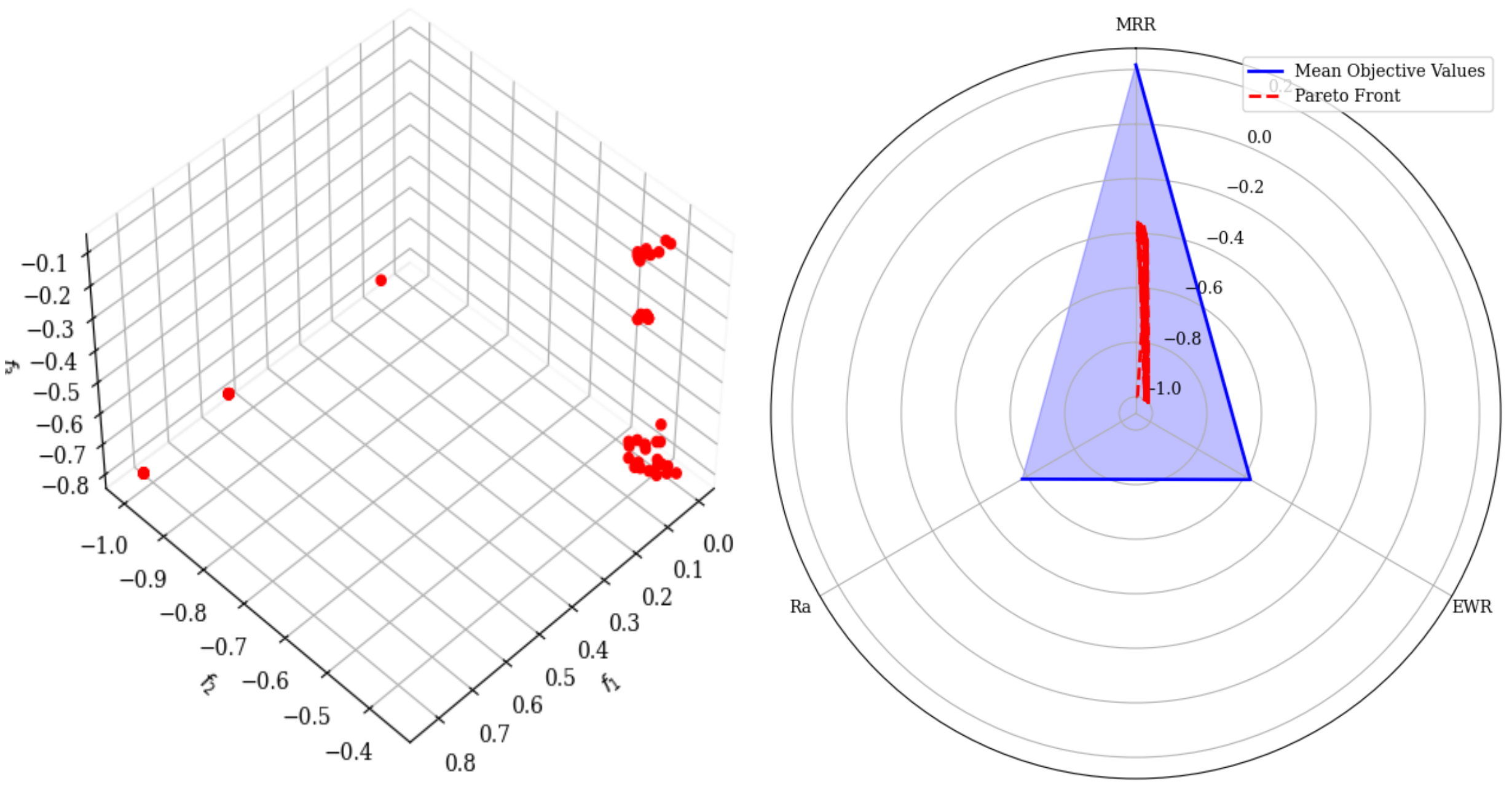

Figure 23. C-TAEA Optimization Method on AdaBoost Prediction Results, (a) Pareto Front Solution, and (b) Pareto Front for Reference Direction in Radar Chart.

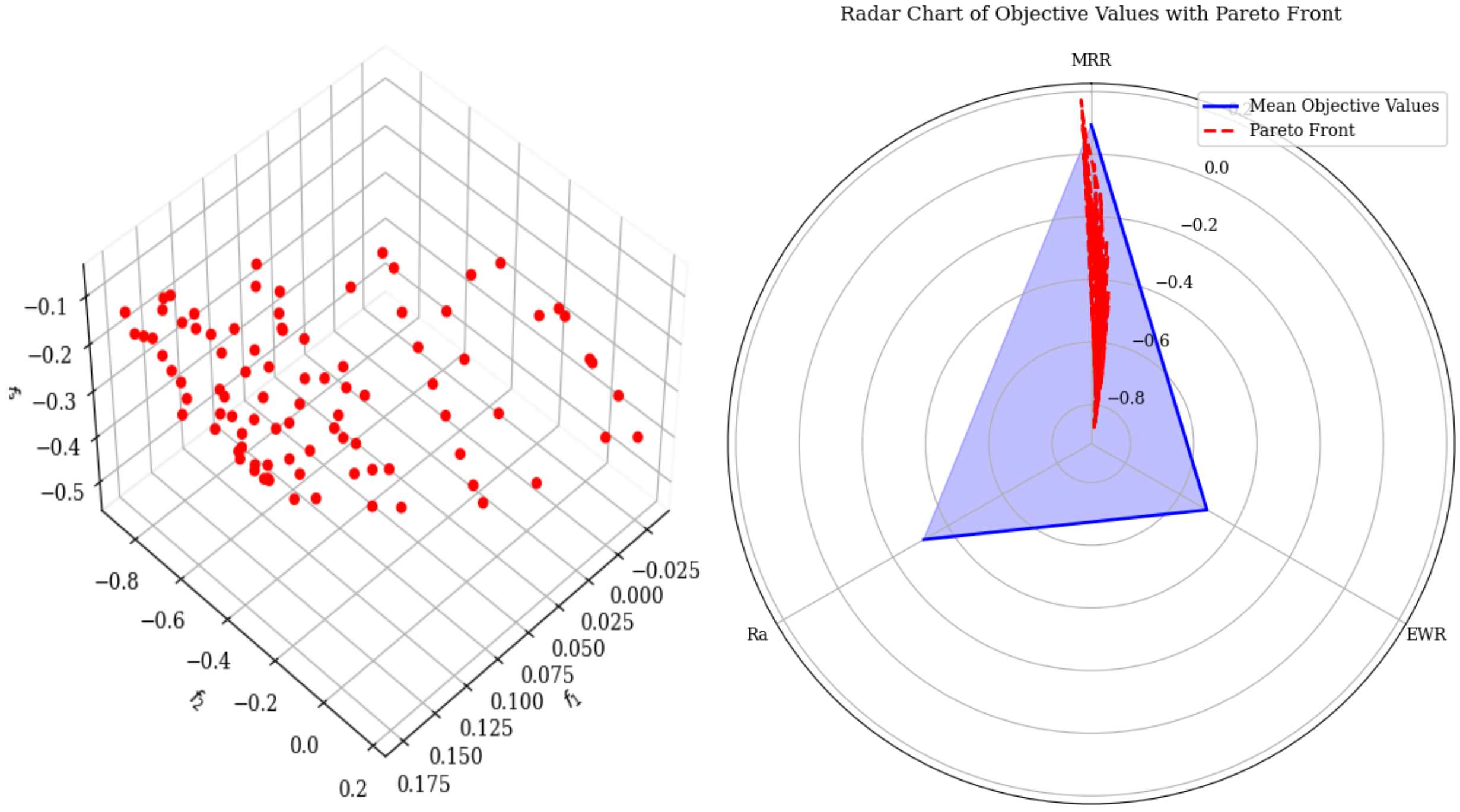

Figure 24. C-TAEA Optimization Method on ElasticNet Prediction Results, (a) Pareto Front Solution, and (b) Pareto Front for Reference Direction in Radar Chart.

# 4. Conclusion

In conclusion, our study underscores the pivotal role of machine learning in enhancing the efficiency and precision of Electrical Discharge Machining (EDM), a complex process vital for precision manufacturing applications. By leveraging ML techniques, specifically evaluating four top-performing models - DNN, XGBoost, AdaBoost, and ElasticNet - against a backdrop of 16 ML models, we achieved comprehensive insights into their predictive capabilities. Through meticulous evaluation across various metrics including Accuracy ($R2$), Mean Squared Error (MSE), Root Mean Squared Error (RMSE), and Mean Absolute Error (MAE), we focused on estimating critical outputs such as Material Removal Rate (MRR), Electrode Wear Rate (EWR), and Surface Roughness (Ra) in Powder-Mixed Electrical Discharge Machining (PMEDM), integrated with a vibration system.

Our evaluations, depicted in Figures 9 and 10, highlighted XGBoost's superior accuracy, with an $R2$ value of 88%, followed by AdaBoost (84%), DNN (82%), and ElasticNet (55%). Radar chart visualization further accentuated the relative performance, with DNN, XGBoost, and AdaBoost exhibiting exceptional performance at 88% accuracy compared to ElasticNet's 55%. This comparative analysis elucidated the strengths and weaknesses of each model, guiding optimization efforts for enhanced predictive capabilities. Furthermore, incorporating Multi-Objective Evolutionary Algorithms (MOEAs) such as NSGA-II, NSGA-III, UNSGA-III, and C-TAEA allowed us to explore and optimize the Pareto front effectively. Notably, the best solutions obtained under NSGA-III for each model showcased promising results, with MRR, EWR, and Ra values as follows: for DNN [11.63, 0.16, 0.22], XGBoost [1.05, 0.60, 0.86], AdaBoost [2.28, 0.37, 0.45], and ElasticNet [1.17, 0.49, 0.71]. Additionally, the similarities observed in the solutions obtained under different MOEAs underscored the potential of UNSGA-III and C-TAEA in achieving comparable performance to established algorithms.

A finely tuned EDM process relies on several factors: the chosen process parameters, the material properties pre- and post-processing, as well as post-processing techniques. These elements collectively contribute to the ex-situ assessment of production quality, gauged through parameters such as EWR and Ra. Concurrently, in-situ monitoring evaluates Material Removal Rate (MRR), another crucial aspect impacting process performance. Empirical evidence suggests that achieving higher surface integrity correlates with MRR being closer to its optimal point within the finishing range, while keeping EWR and Ra at efficient levels for EDM. Leveraging machine learning and MOEAs has significantly advanced the analysis and decision-making processes within EDM operations.

Our findings emphasize the transformative impact of ML and optimization techniques in advancing EDM processes, offering cost-effective, time-efficient, and reliable solutions for precision manufacturing. The elucidation of optimal solutions through MOEAs further contributes to informed decision-making and paves the way for future advancements in PMEDM and related domains.